\documentclass[a4paper,12pt]{extarticle}

\usepackage[english]{babel}
\usepackage[utf8]{inputenc}
\usepackage{amsmath}
\usepackage{amsfonts}
\usepackage{array}
\usepackage{bbold}
\usepackage{hyperref}
\usepackage{amssymb}
\usepackage[top=2.5cm,bottom=2.5cm,left=2.5cm,right=2.5cm]{geometry}

\hypersetup{
colorlinks=false,
linkbordercolor={1 1 1},
citebordercolor={0 0 1}
}

\newcommand{\fld}[3]{\mathcal{C}^{#1}\left(#2\middle|#3\right)}
\newcommand{\p}[1]{\dot{#1}}
\newcommand{\ob}[1]{\overline{#1}}
\newcommand{\ot}[1]{\widetilde{#1}}
\newcommand{\bb}[1]{\mathbb{#1}}
\newcommand{\rk}{\mathbf{r}}

\newcommand{\D}{\mathrm{d}}
\newcommand{\ind}[1]{\mathbf{\mathsf{#1}}}
\newcommand{\oind}[1]{\ob{\mathbf{\mathsf{#1}}}}
\newcommand{\bras}[2]{\langle #1\,#2\rangle}
\newcommand{\braq}[2]{\big[ #1\,#2\big]}

\newcommand{\sgn}{\mathrm{sign}}
\newcommand{\got}[1]{\mathfrak{#1}}
\newcommand{\pgot}[1]{\dot{\mathfrak{#1}}}
\newcommand{\op}[1]{\widehat{#1}}
\newcommand{\Mink}[1]{\mathcal{M}_{#1}}
\newcommand{\be}{\begin{equation}}
\newcommand{\ee}{\end{equation}}

\begin{document}

\begin{flushright}
\vspace{1mm}
FIAN/TD/16-2018\\
\end{flushright}

\vskip 1.5cm

 \begin{center}
 {\large\bf
Scattering amplitudes as multi-particle higher-spin charges \\in the correspondence space}
 \vglue 0.6  true cm

\vskip0.5cm

Y.~O.~Goncharov and M.~A.~Vasiliev

 \vglue 0.3  true cm

{\it I.E.Tamm Department of Theoretical Physics, Lebedev Physical
Institute,\\
Leninsky prospect 53, 119991, Moscow, Russia}

 \end{center}

\begin{center}
\vspace{0.5cm}
goncharov@lpi.ru, vasiliev@lpi.ru\\

\par\end{center}

\numberwithin{equation}{section}

\begin{abstract}
\noindent
Following the  proposal of \cite{higher_rank},
multi-particle scattering amplitudes are represented as conserved higher-spin charges.
 The advantage of such reformulation is that multi-particle  amplitudes acquire the form of an integral of a closed form in the correspondence unifying usual space-time with the twistor-like spinor space. This allows one to  identify seemingly different formulae for amplitudes in terms of twistor and space-time integrals.
 Example of tree MHV amplitudes of Yang-Mills theory is considered in detail. Our results are derived using  unfolded dynamics formulation of massless fields. In these terms all information on the amplitude is contained in a single function $\eta$ that,  even for lower spin amplitudes, can be interpreted as a higher-spin symmetry parameter. The proposed technique can be useful to relate different approaches to amplitude calculations.
\end{abstract}

\section{Introduction}

The celebrated Parke-Taylor formula for the $n$-particle color-ordered tree-level maximally-helicity-violating (MHV) amplitude  with $i$-th and $j$-th gluons carrying negative helicity has the form \cite{PT,PT_proof}
\begin{equation}\label{eq:PT}
\mathcal{A}\left(c_1^+ \dots c_i^-\dots c_j^-\dots c_n^+\right) \propto \mathrm{Tr}\left(T^{c_1}\dots T^{c_n}\right) \, \mathcal{A}^{(ij,1\dots i-1\;i+1\dots j-1\;j+1\dots n)}_{\lambda,\ot{\lambda}}\,,
\end{equation}
where
\begin{equation}\label{eq:PTf}
\mathcal{A}^{(ij,1\dots i-1\;i+1\dots j-1\;j+1\dots n)}_{\lambda,\ot{\lambda}} = \dfrac{\langle{\lambda_i\lambda_j\rangle}^4}{\langle\lambda_1\lambda_2\rangle \langle\lambda_2\lambda_3\rangle \dots \langle\lambda_n\lambda_1\rangle}\,\delta^{(4)}\big(\sum_{I=1}^n \lambda_{Ia} \ot{\lambda}_{I\p{a}}\big).
\end{equation}
Here $\lambda_a,\ot{\lambda}_{\p{a}}$ with $a,\p{a} = 1,2$ are two-component spinors representing light-like momenta $p_{a\p{a}} = \lambda_a\ot{\lambda}_{\p{a}}$, and $\langle\lambda_r\lambda_s\rangle = \varepsilon^{ab}\,\lambda_{ra}\lambda_{sb}$, $\varepsilon^{ab} = -\varepsilon^{ba}$, $\varepsilon^{12} = 1$. Formula \eqref{eq:PTf} acquires  deep interpretation in terms of the underlying twistor and Grassmannian geometry \cite{Witten,BCFW,Hamed_S,Hamed_gras} and led to the discovery of dual superconformal invariance of scattering amplitudes \cite{Dual_Super_Conf,Dual_Super_Conf_2} together with the infinite-dimensional Yangian symmetry for planar maximally supersymmetric Yang-Mills theories  \cite{Yangian}.  Grassmannian integration approach was also applied to calculation of form-factors and off-shell amplitudes \cite{Bork_ff,form_fact,Bork_off_shell}. Interesting \textit{perturbiner} approach of \cite{perturbiner,perturbiner_grav} interpret full Yang-Mills theory as a perturbation to its self-dual (SD) sector. Generalization to the interacting conformal higher-spin (HS) fields  was proposed in \cite{Mc_actions,Mc} within the twistor framework.

Amplitudes operate with free asymptotic states, {\it i.e.}, free particles. It is well known that free fields  exhibit HS symmetries (see {\it e.g.}, \cite{Shaynkman:2001ip,Vasiliev:2001zy}). These are generated by HS conserved charges $\mathcal{Q}_{\eta}$  bilinear in dynamical fields $\mathcal{C}$ \cite{theta} and represented by integrals of an on-shell closed differential form  $\Omega_{\eta} (\mathcal{C})$
\begin{equation}\label{eq:Q}
\mathcal{Q}_{\eta}=\int_{\mathbf{S}} \Omega_{\eta}(\mathcal{C})
\end{equation}
over a cycle $\mathbf{S}$ in the correspondence space $\mathbf{Cor}$ unifying a twistor-like space with spinorial local coordinates $y^{a}$, $\ot{y}^{\p{a}}$ and $4d$-Minkowski space-time with coordinates $x^{a \p{a}}$ \cite{BRST}. Here parameters $\eta(y,\ot{y},x)$ parameterize charges associated with various (lower- and higher-spin) symmetries of the free fields $\mathcal{C}$. That the form $\Omega_{\eta}$  is closed in the correspondence space implies the conservation condition: $\mathcal{Q}_{\eta}$ is independent of local variations of  $\mathbf{S}$. In particular, this allows one to compute HS charges  in the form of integrals over either twistor-like space with spinorial local coordinates $y^{a}$, $\ot{y}^{\p{a}}$ or Minkowski space.

Being bilinear in the fields $\mathcal{C}$ usual HS currents describe only two particles and thus are of little interest for scattering amplitudes. However, as shown in \cite{twistors}, the construction of conserved currents admits a generalization to the multi-particle case with any number of scattering particles. Such currents were argued in \cite{twistors} to form generators of the so-called multi-particle symmetries \cite{multiparticle} underlying multi-particle extensions of HS theories considered in \cite{Coxeter}. The aim of this paper is to reformulate scattering amplitudes as multi-particle charges of free fields. The benefit of such reformulation is that it gives an opportunity to formulate scattering amplitudes both as integrals over spinor (twistor) space and as space-time integrals.  Note that $n$-particle currents are not local in the standard sense being defined in the appropriately extended generalized space containing Cartesian product of $n$ two-dimensional SD planes\footnote{By a SD two-dimensional plane in the complexified Minkowski space we mean linear parametrizaton of $x^{a\p{a}}$ by two variables $\ot{z}^{\p{a}}$ such that $\D x_{c\p{a}}\left(\ot{z}\right)\wedge\D x^{c}{}_{\p{b}}\left(\ot{z}\right) = 0$ ({\it cf.} Section \ref{Sec_Mink}).} in the complexified Minkowski space (see Section \ref{Sec_Mink}). In other words, multi-particle charges contain multiple space-time integrations.

 Let us stress that our proposal is purely kinematical: construction of conserved charges contains functional ambiguity in the parameters $\eta$ which cannot be fixed at the level of free asymptotic states, giving enough freedom to reproduce known amplitudes in the HS framework. The freedom in $\eta$ is to be fixed by the dynamical computation in a theory in question, which is not the aim of this paper. Instead we say that, once $\eta$ is given,  representation (\ref{eq:Q})
   can be used to pass from the representation of the amplitude in the form of twistor integral to that in the form of space-time integral and vice versa. In this paper we identify  parameter $\eta$ associated with tree MHV multi-particle amplitudes. Well known from the HS side (see \textit{e.g.} \cite{twistors,chargesAdS}), this approach seems to be novel in application to amplitudes were it can be used to explain equivalence of seemingly unrelated techniques.

The rest of the paper is organized as follows. In section
\ref{Sec_HS} we recall the construction of unfolded  equations for free HS fields and their tensor products as well as the construction of conserved charges. In Section \ref{Sec_Amp} we represent MHV amplitudes \eqref{eq:PTf} (also for HS fields \cite{BC_hs,Hamed_hs}) as conserved charges and derive form-factors for gauge-invariant HS field strengths. In Section \ref{Sec_Mink} we derive MHV amplitudes in the framework of  generalized Minkowski space \cite{higher_rank} where polylocal products of HS fields live. We conclude in Section \ref{Sec_Conclusion} by  discussing general properties of the relation between scattering amplitudes and conserved multi-particle HS charges as well as some perspectives.

\section{Free higher-spin fields in the  unfolded formalism}\label{Sec_HS}
\subsection{Field equations and their solutions}\label{Sec_1_1}
We start with the rank-one unfolded equation \cite{Vasiliev:1988sa,vasiliev_starprod_ads}
\begin{equation}\label{eq:unfolded_field}
\left(\dfrac{\partial}{\partial x^{a\p{a}}} +
i\dfrac{\partial^2}{\partial y^a\partial \ot{y}^{\p{a}}}\right)\fld{1}{y,\ot{y}}{x} = 0,
\end{equation}
where $x^{a\p{a}}$ with spinorial indices $a,\p{a} = 1,2$ parametrize $4d$ Minkowski space-time which we denote $\Mink{2}$, while $y^a$, $\ot{y}^{\p{a}}$ are auxiliary variables. For $\fld{1}{y,\ot{y}}{x}$ being a formal power series in $y^{a},\,\ot{y}^{\p{a}}$, equation \eqref{eq:unfolded_field} describes dynamics of gauge-invariant field strengths for massless fields of all spins including spins $0$, $1/2$ and $1$ in $4d$ Minkowski space. 
%Various components of the fields $\fld{1}{y,\ot{y}}{x}$ with respect to auxiliary %$y$-variables encode fields of
%different spins along with all their on-shell derivatives. 
The dependence on auxiliary variables in $\fld{1}{y,\ot{y}}{x}$ can be resolved with the aid of $\sigma_-$-cohomology analysis, leaving one with primary multi-spinor fields $c\left(x\right), c_{a}\left(x\right), c_{\p{a}}\left(x\right),\dots,c_{a(2s)},c_{\p{a}(2s)},\dots$, representing (anti-)self-dual field strengths for spins $0,1/2,\dots,s,\dots$  and descendants associated with all their on-shell nontrivial derivatives (see e.g. \cite{vasiliev_starprod_ads}).

Upon Fourier transform with respect to variables $y^a,\ot{y}^{\p{a}}$, equation \eqref{eq:unfolded_field} imposes light-like-momentum constraint
\begin{equation}\label{eq:light_momentum}
p_{a\p{a}}=i\dfrac{\partial}{\partial x^{a\p{a}}} \sim \mu_a \ot{\mu}_{\p{a}}
\end{equation}
familiar in the spinor-helicity formalism (see e.g. \cite{Witten}) and in the hyper-space approach to HS theory (see \cite{Sorokin_hyperspace} for a review). To see that constraint \eqref{eq:light_momentum} indeed implies that $p_{a\p{a}}$ is light-like it suffices to observe that $p^2=\det \big( p_{a\p{a}}\big)$.

In fact, massless fields of spins $s = 1,3/2,2,\dots$ are described by one-forms (gauge potentials)
\begin{equation}\label{eq:gauge_potentials}
\qquad\mathcal{W}\big(y,\ot{y}\big| x\big) = \sum_{m,n = 0}^{\infty}\,
\frac{1}{m!n!} W_{a_1\dots a_m,\p{a}_1\dots \p{a}_n}(x) y^{a_1}\dots y^{a_m}\,
\ot{y}^{\p{a}_1}\dots \ot{y}^{\p{a}_n},
\end{equation}
where $W_{a_1\dots a_m,\p{a}_1\dots\p{a}_n} = \D x^{c\p{c}}\,w_{c\vert a_1\dots a_m,\p{c}\vert\p{a}_1\dots\p{a}_n}$. Spin-$s$
 fields have $m+n = 2(s-1)$. The familiar lower-spin examples include electromagnetic potential $W$ for $s=1$ and vierbein $W_{a\p{a}}$ with Lorentz connection $W_{ab}$, $W_{\p{a}\p{b}}$ for $s=2$. At the free-field level zero-forms $\fld{1}{y,\ot{y}}{x}$ that satisfy equation \eqref{eq:unfolded_field}  are related to HS potentials \eqref{eq:gauge_potentials} via  relation \cite{Vasiliev:1988sa,vasiliev_starprod_ads}
\begin{equation}\label{eq:comst}
\mathcal{DW} = H^{ab}\dfrac{\partial^2}{\partial y^a \partial y^b} \,\fld{1}{y,0}{x} + \ot{H}^{\p{a}\p{b}}\dfrac{\partial^2}{\partial \ot{y}^{\p{a}}\partial \ot{y}^{\p{b}}}\,\fld{1}{0,\ot{y}}{x}
\end{equation}
often referred to as Central On-Mass-Shell Theorem. Here $H^{ab} = \D x^{a}{}_{\p{c}}\wedge \D x^{b\p{c}}$, $\ot{H}^{\p{a}\p{b}} = \D x_{c}{}^{\p{a}}\wedge \D x^{c\p{b}}$ for the Cartesian coordinates of the flat Minkowski space and $\mathcal{D}$ is a nilpotent covariant differential, $\mathcal{D}^2 = 0$. Spinorial indices are raised and lowered by the $SL(2)$-invariant antisymmetric tensors $\varepsilon_{ab}$, $\varepsilon^{ab}$ ($\varepsilon_{12} = \varepsilon^{12} = 1$) according to the rules $P^{a} = \varepsilon^{ab}\, P_{b}$ and $Q_{a} = Q^b\,\varepsilon_{ba}$ implying $\bras{P}{Q}=\varepsilon_{ab}P^a Q^b = \varepsilon^{ab}P_a Q_b = P_c\,Q^c$ (the rules for dotted indices are analogous with the bracket denoted by $\braq{\ot{P}}{\ot{Q}}$). In the spin-$s$ sector, \eqref{eq:comst} is equivalent to the following set of differential equations: % \cite{Vasiliev_freefields}
\begin{equation}\label{eq:comst_1}
    \D W = H^{aa}\,c_{aa} + \ot{H}^{\p{a}\p{a}}\,\ot{c}_{\p{a}\p{a}}
\end{equation}
 for $s = 1$ and
\begin{multline}\label{eq:comst_s}
\D W_{a(m),\p{a}(n)} + m\,\theta\left(n-m\right)\D x_{a\p{c}}
\wedge W_{a(m-1),}{}^{\p{c}}{}_{\p{a}(n)} +\\ n\,\theta\left(m-n\right)\D x_{c\p{a}}
\wedge W^{c}{}_{a(m),}{}_{\p{a}(n-1)} = 0 \quad \text{for}\quad m+n = 2\left(s-1\right),
\quad m,n>0,\\
 \D W_{a\left(2(s-1)\right)} = H^{aa}\,c_{a(2s)},\quad \D W_{\p{a}\left(2(s-1)\right)} = \ot{H}^{\p{a}\p{a}}\,\ot{c}_{\p{a}\left(2s\right)}\,.\qquad\qquad
\end{multline}
for $s\ge 3/2$. Here $\theta\left(n\right) = 1(0)$ for $n\ge 0(n<0)$. Indices denoted by the same letter are symmetrized.

Gauge transformations $\delta \mathcal{W} = \mathcal{D}\epsilon$ with an arbitrary zero-form gauge parameter
\begin{equation}\label{eq:gauge_transform}
\epsilon = \sum_{m,n = 0}^{\infty}\, \dfrac{1}{m!n!}\epsilon_{a_1\dots a_m,\p{a}_1\dots \p{a}_n}(x) y^{a_1}\dots y^{a_m}\,\ot{y}^{\p{a}_1}\dots \ot{y}^{\p{a}_n}
\end{equation}
leave  \eqref{eq:comst}  (equivalently, \eqref{eq:comst_1}, \eqref{eq:comst_s}) invariant. Equations \eqref{eq:comst_1}, \eqref{eq:comst_s} thus define the spin-$s$ gauge invariant zero-forms $c_{a(2s)}$ and $c_{\dot a (2s)}$  via the $\lfloor s\rfloor$-th derivative of the
field $W_{a(m),\p{a}(n)}$ with $| m-n| \le 1$, where $\lfloor s \rfloor$ is the integer part of $s$.

As is well known, gauge potentials \eqref{eq:gauge_potentials} are needed for locality of field interactions. Therefore one-forms \eqref{eq:gauge_potentials} rather than zero-forms \eqref{eq:unfolded_field} serve as the main ingredient for calculation of scattering amplitudes. However in this paper it is more convenient to establish first the correspondence between HS charges and scattering amplitudes in terms of zero-forms. It should be stressed that since amplitudes themselves are not local, the two languages are equivalent. Relation to one-forms \eqref{eq:gauge_potentials} will be discussed in Section \ref{Sec_FormFactors}.

As shown in \cite{Gelfond:2003vh} equation \eqref{eq:unfolded_field} can be generalized to  tensor powers of HS fields by introducing $\rk\ge 2$ copies of auxiliary variables $y^{a}_I$, $\ot{y}^{\p{a}}_J$ (where $I,J=1...\rk$) with  rank-$\rk$ fields $\fld{\rk}{y,\ot{y}}{x}$  obeying rank-$\rk$ unfolded equation
\begin{equation}\label{eq:unfolded}
\left(\dfrac{\partial}{\partial x^{a\p{a}}} + i\,\delta_{IJ}\dfrac{\partial^2}{\partial y^a_I\partial \ot{y}^{\p{a}}_J}\right)\fld{\rk}{y,\ot{y}}{x} = 0.
\end{equation}
Products of $\rk$ solutions to \eqref{eq:unfolded_field} $\prod_{I=1}^{\rk}\fld{1}{y_I,\ot{y}_I}{x}$ obviously solve \eqref{eq:unfolded}. The rank-two field $\fld{2}{y,\ot{y}}{x}$ obeying \eqref{eq:unfolded} encodes HS conserved currents in $4d$ Minkowski space \cite{Gelfond:2003vh}. Rank-$\rk>2$ fields  (also referred to as \textit{higher-rank currents}) were introduced in \cite{Gelfond:2003vh,higher_rank}.

Solutions $\fld{\rk}{y,\ot{y}}{x}$ to \eqref{eq:unfolded} can be expanded into plane waves
\begin{equation}
\chi_{\mu,\ot{\mu}} \sim \exp\left[i\big(\mu_{Ia}\ot{\mu}_{I\p{a}}\,x^{a\p{a}} + \mu_{Ia}\,y_{I}{}^a + \ot{\mu}_{I\p{a}}\,\ot{y}_{I}{}^{\p{a}}\big)\right].
\end{equation}

Following  \cite{theta} equation \eqref{eq:unfolded} can be transformed into a first-order PDE as follows. Consider a partition $\rk = m + \ob{m}$ and represent $\fld{\rk}{y,\ot{y}}{x}$ as a half-Fourier transform
\begin{equation}
\fld{\rk}{y,\ot{y}}{x} = \dfrac{1}{\left(2\pi\right)^{2\rk}}\int \D^{2m} \lambda_{1...m}\,\D^{2\ob{m}}\ot{\lambda}_{\ob{m+1}...\ob{\rk}}\;e^{i\big(y^a_{\ind{i}}\lambda_a^{\ind{i}} + \ot{y}^{\p{a}}_{\ob{\ind{i}}}\ot{\lambda}_{\p{a}}^{\ob{\ind{i}}}\big)}\;g^{(m,\ob{m})}\left(\lambda_{\ind{i}},y_{\oind{i}};\ot{y}_{\ind{j}},\ot{\lambda}_{\oind{j}}\middle| x\right),
\end{equation}
where $\ind{i},\ind{j}=1...m$ and $\ob{\ind{i}},\ob{\ind{j}}=\ob{m+1}...\ob{\rk}$. Equation \eqref{eq:unfolded}  rewritten in terms of Fourier components $g^{(m,\ob{m})}$ has the first-order form
\begin{equation}\label{eq:unfolded_Fourier}
\left(\dfrac{\partial}{\partial x^{a\p{a}}} -\delta_{\ind{i}\ind{j}}\,\lambda_a^{\ind{i}}\dfrac{\partial}{\partial \ot{y}^{\p{a}}_{\ind{j}}}- \delta_{\oind{i}\oind{j}}\,\ot{\lambda}_{\p{a}}^{\oind{i}}\dfrac{\partial}{\partial y^{\p{a}}_{\oind{j}}}\right)g^{(m,\ob{m})}\left(\lambda_{\ind{i}},y_{\oind{i}};\ot{y}_{\ind{j}},\ot{\lambda}_{\ob{\ind{j}}}\middle| x\right) = 0
\end{equation}
 with characteristics
 \begin{equation}\label{eq:characteristics}
  \lambda_{\ind{i}\,a},\; \ot{y}^{\p{a}}_{\ind{i}} + x^{a\p{a}}\lambda_{\ind{i}\,a}\quad
\text{and}\quad\ot{\lambda}_{\oind{j}\,\p{a}},\; y^{a}_{\oind{j}} + x^{a\p{a}}\ot{\lambda}_{\oind{j}\,\p{a}},
 \end{equation}
where $\lambda_{\ind{i}\,a} = \delta_{\ind{i}\ind{j}}\,\lambda_{a}^{\ind{j}}$ and $\ot{\lambda}_{\oind{i}\p{a}} = \delta_{\oind{i}\oind{j}}\,\ot{\lambda}_{\p{a}}^{\oind{j}}$. A useful solution that depends only on spinor variables being free of the  space-time dependence is
\begin{equation}\label{eq:link}
\rho_{\ind{i}\oind{j}} = \lambda_{\ind{i}\,a}y_{\oind{j}}^{a} - \ot{\lambda}_{\oind{j}\,\p{a}} \ot{y}_{\ind{i}}^{\p{a}}.
\end{equation}
Analogous expression was introduced in \cite{Hamed_S} for the so-called ``link representation'' of scattering amplitudes.

It is convenient to introduce the extended space $\mathbf{Cor}_2^{(m,\ob{m})} = \Mink{2}\times \mathbb{R}^{4\rk}$ parameterized by space-time coordinates $x^{a\p{a}}$ together with spinor coordinates $y^a_{\oind{i}}$, $\lambda_{\ind{i}a}$ ($2\rk$ variables) and $\ot{y}^{\p{a}}_{\ind{i}}$, $\ot{\lambda}_{\oind{i}\p{a}}$ ($2\rk$ variables), $\dim \mathbf{Cor}_2^{(m,\ob{m})} = 4 + 4\rk$. On-shell condition \eqref{eq:unfolded_Fourier} implies that functions $g^{(m,\ob{m})}$ on $\mathbf{Cor}_2^{(m,\ob{m})}$ are functions of characteristics \eqref{eq:characteristics}.

We will use plane-wave solutions to \eqref{eq:unfolded_Fourier} of the form
\begin{multline}\label{eq:helicity}
\chi_{\mu,\ot{\mu}}\left(\lambda_{\ind{i}},y_{\ob{\ind{i}}};\ot{y}_{\ind{j}},
\ot{\lambda}_{\ob{\ind{j}}}\middle| x\right) =
\mathrm{exp}\left[i\left(\ot{\mu}^{\ind{i}}_{\p{a}}\left(\ot{y}_{\ind{i}}^{\p{a}} +
x^{a\p{a}}\lambda_{\ind{i}\,a}\right) + \mu^{\oind{j}}_{a}\left(y_{\oind{j}}^{a} +
x^{a\p{a}}\ot{\lambda}_{\oind{j}\,\p{a}}\right)\right)\right]\cdot \\ \prod_{\ind{j},\,a}
\delta\left(\lambda^{\ind{j}}_a - \mu^{\ind{j}}_a\right)\,\prod_{\oind{j},\,\p{a}}\delta
\left(\ot{\lambda}^{\oind{j}}_{\p{a}}-\ot{\mu}^{\oind{j}}_{\p{a}}\right)\,
\end{multline}
with $\mu_{Ia}$ and $\ot{\mu}_{I\p{a}}$  treated as parameters.

\subsection{On-shell closed differential forms}\label{Sec_diff_forms}
Given  solution $g^{(m,\ob{m})}$ to  \eqref{eq:unfolded_Fourier} generates an on-shell closed differential $4\rk$-form in $\mathbf{Cor}_2^{(m,\ob{m})}$
 \cite{theta}
\begin{equation}\label{eq:form}
\Omega^{(m,\ob{m})}\left[g\right] = \D^{2m}\lambda\,\D^{2\ob{m}}\ot{\lambda}
\;\D^{2\ob{m}}\big(y + x\ot{\lambda}\big)\,\D^{2m}\big(\ot{y} + x\lambda\big)\;
g^{(m,\ob{m})}\big(\lambda_{\ind{i}},y_{\ob{\ind{i}}};\ot{y}_{\ind{j}},\ot{\lambda}_{\oind{j}}\big| x\big).
\end{equation}
 $\Omega^{(m,\ob{m})}\left[g\right]$ contains a top form in differentials of characteristics \eqref{eq:characteristics}, while $g^{(m,\ob{m})}$  is some function of the same characteristics  by virtue of \eqref{eq:unfolded_Fourier}. As a result, de Rham differential in  $\mathbf{Cor}_2^{(m,\ob{m})}$,
\begin{equation}\label{eq:deRham}
\D_{\mathbf{Cor}} = \D\lambda_{\ind{i}a}\dfrac{\partial}{\partial \lambda_{\ind{i}a}} + \D\ot{\lambda}_{\oind{j}\p{a}}\dfrac{\partial}{\partial \ot{\lambda}_{\oind{j}\p{a}}} + \D y^{a}_{\oind{i}}\dfrac{\partial}{\partial y^{a}_{\oind{i}}} + \D \ot{y}^{\p{a}}_{\ind{j}}\dfrac{\partial}{\partial \ot{y}^{\p{a}}_{\ind{j}}} + \D x^{a\p{a}}\dfrac{\partial}{\partial x^{a\p{a}}}\,,
\end{equation}
gives zero on \eqref{eq:form},
\begin{equation}\label{eq:DO}
\D_{\mathbf{Cor}} \Omega^{(m,\ob{m})}\left[g\right]=0\,.
\end{equation}
More in detail, since $g^{(m,\ob{m})}$ solves \eqref{eq:unfolded_Fourier}, it is a function of characteristics \eqref{eq:characteristics}. As a result,
\begin{multline}
\D_{\mathbf{Cor}}g^{(m,\ob{m})} = \left(\D\lambda_{\ind{i}a}\bigg(\dfrac{\partial}{\partial \lambda_{\ind{i}a}} - x^{a\p{a}}\dfrac{\partial}{\partial \ot{y}^{\p{a}}_{\ind{i}}}\bigg) + \D\ot{\lambda}_{\oind{j}\p{a}}\bigg(\dfrac{\partial}{\partial \ot{\lambda}_{\oind{j}\p{a}}} - x^{a\p{a}}\dfrac{\partial}{\partial y^{a}_{\oind{j}}}\bigg) \right.+ \\ \left. \D\big(y^{a}_{\oind{j}} + x^{a\p{a}}\,\ot{\lambda}_{\oind{j}\p{a}}\big)\dfrac{\partial}{\partial y^{a}_{\oind{j}}} + \D\big(\ot{y}^{\p{a}}_{\ind{i}} + x^{a\p{a}}\,\lambda_{\ind{i}\p{a}}\big)\dfrac{\partial}{\partial \ot{y}^{\p{a}}_{\ind{i}}}\right)\,g^{(m,\ob{m})},
\end{multline}
\textit{i.e.} $\D_{\mathbf{Cor}}g^{(m,\ob{m})}$ is proportional to differentials of characteristics \eqref{eq:characteristics}. Since $\Omega^{(m,\ob{m})}\left[g\right]$ \eqref{eq:form} already contains a top form in characteristics \eqref{eq:characteristics}, \eqref{eq:DO} is true.

We  focus on the plane-wave  configuration \eqref{eq:helicity}  considering $\Omega^{(m,\ob{m})}$ of the form
\begin{equation}\label{eq:form_helicity}
\Omega^{(m,\ot{m})}_{\mu,\ob{\mu}}\left[\eta\right] = \D^{2m}\lambda\,\D^{2\ob{m}}\ot{\lambda}\;\D^{2\ob{m}}\big(y + x\ot{\lambda}\big)\,\D^{2m}\big(\ot{y} + x\lambda\big)\;\eta\,\chi_{\mu,\ot{\mu}},
\end{equation}
where $\eta= \eta \big(\lambda_{\ind{i}},y_{\ob{\ind{i}}};\ot{y}_{\ind{j}}, \ot{\lambda}_{\oind{j}}\big| x\big)$ is a solution to \eqref{eq:unfolded_Fourier}. Note that since \eqref{eq:unfolded_Fourier} is a first-order PDE, the product of two its solutions $\eta$ and $\chi_{\mu,\ot{\mu}}$ is again a solution. It should be stressed that on-shell closed differential form \eqref{eq:form_helicity} has vast freedom  due to the functional parameter $\eta$ which is also allowed to depend on the plane wave parameters $\mu_{Ia}$, $\ot{\mu}_{I\p{a}}$ \eqref{eq:helicity}. On the other hand, to preserve translational invariance $\eta$ is demanded to be free of explicit $x$-dependence. This is achieved for $\eta$ depending only on $\lambda_{\ind{i}a}$, $\ot{\lambda}_{\oind{j}\p{a}}$ and $\rho_{\ind{i}\oind{j}}$ \eqref{eq:link}
\begin{equation}\label{eq:eta_uniform}
\eta = \eta (\lambda,\ot{\lambda}, \rho)\,.
\end{equation}
Possible dependence on the plane-wave parameters $\mu_{Ia}$, $\ot{\mu}_{I\p{a}}$ will be implicit.

Integration of \eqref{eq:form_helicity}  over a $4\rk$-dimensional hypersurface $\mathbf{S}\subset\mathbf{Cor}_2^{(m,\ob{m})}$ gives conserved charge
\begin{equation}\label{eq:amplitude}
\mathcal{Q}^{(m,\ob{m})}_{\mu,\ot{\mu}}\left[\eta\right]:=\int_{\mathbf{S}} \Omega^{(m,\ob{m})}_{\mu,\ot{\mu}}\left[\eta\right]
\end{equation}
independent of local variations of $\mathbf{S}$. Our goal is to express multi-particle amplitudes in the form of such charges.

\section{MHV amplitudes}\label{Sec_Amp}

\subsection{3-particle amplitude}

As mentioned in  Introduction, the $n$-particle color-ordered tree amplitude is given by Parke-Taylor formula \eqref{eq:PT}. Our aim is to represent \eqref{eq:PTf} in the form \eqref{eq:amplitude}. For $3$-particle amplitude with negative-helicity gluons  $1$ and $2$ expression \eqref{eq:PTf} gives
\begin{equation}\label{eq:PTf_3}
\mathcal{A}^{(12,3)}_{\mu,\ot{\mu}} = \dfrac{\langle{\mu_1\mu_2\rangle}^3}{\langle\mu_2\mu_3\rangle \langle\mu_3\mu_1\rangle}\,\delta^{(4)}\big(\sum_{I=1}^3 \mu_{Ia} \ot{\mu}_{I\p{a}}\big).
\end{equation}
To represent this expression in the form \eqref{eq:amplitude} we set  $\rk = 3$ with $m=2$ and $\ob{m} = 1$ and $\eta = \eta^{(12,\ob{3})} := \prod_{\ind{i}=1,2,\oind{j}=\ob{3}}\mathrm{sign} \left(\rho_{\ind{i}\oind{j}}\right)$ with $\rho_{\ind{i}\oind{j}}$  \eqref{eq:link}  ($\mathrm{sign}\left(x\right)$ is the sign function). For the integration surface $\mathbf{S}_1\subset \mathbf{Cor}_2^{(2,1)}$ such that $x=0$, using \eqref{eq:helicity} and \eqref{eq:form_helicity} we obtain
\begin{multline}\label{eq:amplitude_3pt}
\mathcal{Q}^{(2,1)}_{\mu,\ot{\mu}}\left[\eta^{(12,\ob{3})}\right] = \int \D^{4}\lambda\,\D^2\ot{\lambda}\,\D^2 y\,\D^4\ot{y}\;\mathrm{exp}\left(i\ot{\mu}_{\ind{i}}\ot{y}_{\ind{i}} + i\mu_{\oind{j}}y_{\oind{j}}\right)\\ \delta^{(2)}\big(\lambda_{1}-\mu_{1}\big)\,\delta^{(2)}\big(\lambda_{2}-\mu_{2}\big)\,\delta^{(2)}\big(\ot{\lambda}_{\ob{3}}-\ot{\mu}_{\ob{3}}\big)\,\sgn\left(\rho_{1\ob{3}}\right)\,\sgn\left(\rho_{2\ob{3}}\right).
\end{multline}
Integrating out delta-functions this gives
\begin{equation}\label{eq:MHV3_raw}
\mathcal{Q}^{(2,1)}_{\mu,\ot{\mu}}\left[\eta^{(12,\ob{3})}\right] = \int \D^2 y\,\D^4\ot{y}\;\mathrm{exp}\left(i\ot{\mu}_{\ind{i}}\ot{y}_{\ind{i}} + i\mu_{\oind{j}}y_{\oind{j}}\right)\,\sgn\big(\mu_{1}y_{\ob{3}} - \ot{\mu}_{\ob{3}}\ot{y}_{1}\big)\, \sgn\big(\mu_{2}y_{\ob{3}} - \ot{\mu}_{\ob{3}}\ot{y}_{2}\big).
\end{equation}

To bring this expression the form of MHV amplitude we  change
integration variables. Consider expression \eqref{eq:link}
\begin{equation}\label{eq:change_variables}
\mu_{\ind{i}\,a} y_{\oind{j}}^{a} - \ot{\mu}_{\oind{j}\,\p{a}} \ot{y}_{\ind{i}}^{\p{a}} = \rho_{\ind{i}\oind{j}}
\end{equation}
as a system of linear equations allowing us to express $y_{\ob{3}}^a$ and $\ot{y}_{1,2}^{\p{a}}$ via variables $\rho_{1\ob{3}}$, $\rho_{2\ob{3}}$ and $\ot{\xi}_{\ind{i}}^{\p{a}}$ as follows
\begin{equation}\label{eq:reparametrization}
    y^a_{\oind{j}} = \dfrac{\varepsilon_{\ind{ik}}\varepsilon^{ac}\,\mu_{\ind{k}c}}{\bras{\mu_1}{\mu_2}}\,\rho_{\ind{i}\oind{j}} + \dfrac{\varepsilon_{\ind{ik}}\varepsilon^{ac}\,\mu_{\ind{k}c}\,\ot{\mu}_{\oind{j}\p{a}}}{\bras{\mu_1}{\mu_2}}\,\ot{\xi}_{\ind{i}}^{\p{a}},\quad \ot{y}_{\ind{i}}^{\p{a}} = \ot{\xi}_{\ind{i}}^{\p{a}},
\end{equation}
where $\varepsilon_{\ind{ik}} = -\varepsilon_{\ind{ki}}$ and $\varepsilon_{12} = 1$. This allows us to rewrite $\ot{\mu}_{\ind{i}}\ot{y}_{\ind{i}} + \mu_{\oind{j}}y_{\oind{j}}$ in the exponential of \eqref{eq:MHV3_raw} in the  form
\begin{equation}
\ot{\mu}_{\ind{i}}\ot{y}_{\ind{i}} + \mu_{\oind{j}}y_{\oind{j}} = \rho_{\ind{i}\oind{j}} \,\dfrac{\langle\mu_{\oind{j}} \,\varepsilon_{\ind{i}\ind{k}}\mu_{\ind{k}}\rangle}{\langle \mu_1 \mu_2\rangle} + \ot{\xi}_{\ind{i}}^{\p{a}}\left(\ot{\mu}_{\ind{i}\p{a}} + \dfrac{\langle\mu_{\oind{j}} \,\varepsilon_{\ind{i}\ind{k}}\mu_{\ind{k}}\rangle}{\langle \mu_1 \mu_2\rangle}\cdot \ot{\mu}_{\oind{j}\p{a}}\right).
\end{equation}
It is also straightforward to calculate the Jacobian
\begin{equation}\label{eq:jacobian_3}
\D^2 y\,\D^4\ot{y} = \dfrac{1}{\langle \mu_1\mu_2\rangle}\,\D^2\rho\, \D^4\ot{\xi}\,.
\end{equation}
As a result, one arrives at
\begin{multline}\label{eq:MHV3_raw2}
\mathcal{Q}^{(2,1)}_{\mu,\ot{\mu}}\left[\eta^{(12,\ob{3})}\right] =
\\ \dfrac{1}{\langle \mu_1\mu_2\rangle}\,\int \D^2\rho\, \D^4\ot{\xi}\;
\mathrm{exp}\left[i\,\rho_{\ind{i}\oind{j}} \,\dfrac{\langle\mu_{\oind{j}} \,
\varepsilon_{\ind{i}\ind{k}}\mu_{\ind{k}}\rangle}{\langle \mu_1 \mu_2\rangle}\right]\,
\mathrm{exp}\left[i \,\ot{\xi}_{\ind{i}}^{\p{a}}
\left(\ot{\mu}_{\ind{i}\p{a}} +
\dfrac{\langle\mu_{\oind{j}} \,\varepsilon_{\ind{i}\ind{k}}\mu_{\ind{k}}\rangle}
{\langle \mu_1 \mu_2\rangle}\cdot \ot{\mu}_{\oind{j}\p{a}}\right)\right]\cdot
\prod_{\ind{i},\oind{j}}\sgn\left(\rho_{\ind{i}\oind{j}}\right) = \\ =
\langle \mu_1\mu_2\rangle^5\,\int \D^2\rho\, \D^4\ot{\xi}\;\mathrm{exp}
\left[i\,\rho_{\ind{i}\oind{j}} \,\langle\mu_{\oind{j}} \,
\varepsilon_{\ind{i}\ind{k}}\mu_{\ind{k}}\rangle\right]\,\mathrm{exp}
\big[i \,\ot{\xi}_{\ind{i}}^{\p{a}}F_{\ind{i}\p{a}}\big ]
 \prod_{\ind{i},\oind{j}}\sgn\left(\rho_{\ind{i}\oind{j}}\right) \propto  \dfrac{\langle \mu_1\mu_2\rangle^5}{\langle\mu_2\mu_3\rangle\langle\mu_3\mu_1\rangle}\,\delta^{(4)}\left(F\right),
\end{multline}
where
\begin{equation}
F_{\ind{i}\p{a}} = \langle \mu_1 \mu_2\rangle\,\ot{\mu}_{\ind{i}\p{a}} +
\langle\mu_{\oind{j}} \,\varepsilon_{\ind{i}\ind{k}}\mu_{\ind{k}}\rangle\, \ot{\mu}_{\oind{j}\p{a}}.
\end{equation}
Integration over the variables $\rho$ in \eqref{eq:MHV3_raw2} is performed using the well-known formula \cite{Gelf_Shil}
\begin{equation}\label{eq:formula}
\int_{-\infty}^{+\infty} \,e^{ikx}\,\mathrm{sign}\left(x\right) \D x = \frac{2i}{k}.
\end{equation}
The factor of  $\delta^{(4)}\left(F\right)$ in \eqref{eq:MHV3_raw2} implies momentum conservation $\delta^{(4)}\left(\sum_I\mu_{Ia}\ot{\mu}_{I\p{a}}\right)$. Indeed, denote $P_{a\p{a}} = \mu_{\ind{i}a}\ot{\mu}_{\ind{i}\p{a}} + \mu_{\oind{i}a}\ot{\mu}_{\oind{i}\p{a}}$.
Then
\begin{equation}
\varepsilon^{ab}\,\mu_{\ind{i}a}\,P_{b\p{a}} = \varepsilon_{\ind{i}\ind{j}}\,F_{\ind{j}\p{a}},
\end{equation}
which implies $\delta^4\left(F\right) = \langle \mu_1 \mu_2\rangle^{-2}\,\delta^4\left(P\right)$. The final result
\begin{equation}\label{eq:MHV3_raw3}
\mathcal{Q}^{(2,1)}_{\mu,\ot{\mu}}\left[\eta^{(12,\ob{3})}\right] \propto \dfrac{\langle \mu_1 \mu_2\rangle^3}{\langle\mu_2\mu_3\rangle\langle\mu_3\mu_1\rangle}\,\delta^{(4)}\big(\sum_I\mu_{Ia}\ot{\mu}_{I\p{a}}\big)
\end{equation}
gives \eqref{eq:PTf_3} up to a numerical factor
\begin{equation}\label{eq:result_3pt}
\mathcal{Q}^{(2,1)}_{\mu,\ot{\mu}}\left[\eta^{(12,\ob{3})}\right] \propto \mathcal{A}^{(12,3)}_{\mu,\ot{\mu}}.
\end{equation}

\subsection{MHV $n$-point function}
\label{n-pt}

In the case $\rk = n$ with $m=2$ and $\ob{m} = n-2$, expression \eqref{eq:amplitude} for $\eta = \eta^{(12,\ob{3}\dots\ob{n})}:=\prod_{\ind{i}=1,2,\oind{j}=\ob{3}\dots\ob{n}}
\sgn\big(\rho_{\ind{i}\oind{j}}\big)$ gives
\begin{multline}\label{eq:amplitude_n}
\mathcal{Q}^{(2,n-2)}_{\mu,\ot{\mu}}\left[\eta^{(12,\ob{3}\dots\ob{n})}\right] = \int \D^4\lambda\,\D^{2(n-2)}\ot{\lambda}\,\D^{2(n-2)}y\,\D^{4}\ot{y}\;\mathrm{exp}\left(i\ot{\mu}_{\ind{i}}\ot{y}_{\ind{i}} + i\mu_{\oind{j}}y_{\oind{j}}\right)\\ \delta^{(4)}\big(\lambda-\mu\big)\,\delta^{\left(2(n-2)\right)}\big(\ot{\lambda}-\ot{\mu}\big)\,\prod_{\ind{i},\oind{j}}\sgn\big(\lambda_{\ind{i}a}y^{a}_{\oind{j}} - \ot{\lambda}_{\oind{j}\p{a}}\ot{y}^{\p{a}}_{\ind{i}}\big) = \\ = \int \D^{2(n-2)}y\,\D^{4}\ot{y}\;\mathrm{exp}\left(i\ot{\mu}_{\ind{i}}\ot{y}_{\ind{i}} + i\mu_{\oind{j}}y_{\oind{j}}\right)\,\prod_{\ind{i},\oind{j}}\sgn\left(\mu_{\ind{i}}y_{\oind{j}} - \ot{\mu}_{\oind{j}} \ot{y}_{\ind{i}}\right).
\end{multline}
Analogously to the case of 3-particle amplitude consider the change of variables from $y_{\oind{i}}^a$ ($2(n-2)$ variables) and $\ot{y}_{\ind{i}}^{\p{a}}$ ($4$ variables) to
$\rho_{\ind{i}\oind{j}}$ \eqref{eq:change_variables} ($2(n-2)$ variables) and  $\ot{\xi}_{\ind{i}}^{\p{a}}$ ($4$ variables) resulting from  equations \eqref{eq:change_variables}. Again, this gives parametrization \eqref{eq:reparametrization} and integration measure
\begin{equation}\label{eq:jacobian_n}
\D^{2(n-2)}y\,\D^{4}\ot{y} = \dfrac{1}{\langle\mu_1 \mu_2\rangle^{n-2}}\,\D^{2(n-2)}\rho\,\D^4\ot{\xi}.
\end{equation}
The final result is
\begin{equation}\label{eq:MHVn_pre}
\mathcal{Q}^{(2,n-2)}_{\mu,\ot{\mu}}\left[\eta^{(12,\ob{3}\dots\ob{n})}\right] \propto \dfrac{\langle\mu_1\mu_2\rangle^n}{\prod_{\ind{i}=1,2,\oind{j}=\ob{3}\dots\ob{n}}\langle\mu_{\ind{i}}\mu_{\oind{j}}\rangle}\,\delta^{(4)}\big(\sum_I\mu_{Ia}\ot{\mu}_{I\p{a}}\big).
\end{equation}
  At $n=4$ it reproduces  $4$-particle amplitude with the fixed ordering of gluons, \textit{e.g.} $1 \ob{3} 2\ob{4}$, corresponding to the denominator $\langle\mu_1\mu_{\ob{3}}\rangle \langle\mu_{\ob{3}}\mu_2\rangle \langle\mu_2\mu_{\ob{4}}\rangle \langle\mu_{\ob{4}}\mu_1\rangle$ in \eqref{eq:MHVn_pre}:
\begin{equation}
\mathcal{Q}^{(2,2)}_{\mu,\ot{\mu}}\left[\eta^{(12,\ob{3}\dots\ob{n})}\right] \propto \dfrac{\langle\mu_1\mu_2\rangle^4}{\langle\mu_{1}\mu_{\ob{3}}\rangle \langle\mu_{\ob{3}}\mu_{2}\rangle \langle\mu_{2}\mu_{\ob{4}}\rangle \langle\mu_{\ob{4}}\mu_{1}\rangle}\;\delta^{(4)}\big(\sum_I\mu_{Ia}\ot{\mu}_{I\p{a}}\big).
\end{equation}

Though \eqref{eq:MHVn_pre} does not match $n$-particle MHV amplitude for $n>4$ directly (\textit{cf.}
\eqref{eq:PTf}), using the freedom  in the choice of parameter $\eta$ in \eqref{eq:amplitude} to set
\begin{equation}\label{eq:toMHV_n1}
\eta_{\mu}^{(\mathbf{1})} = \dfrac{\bras{\lambda_1}{\mu_{\ob{3}}}...\bras{\lambda_1}{\mu_{\ob{n}}}\cdot \bras{\lambda_2}{\mu_{\ob{3}}}...\bras{\lambda_2}{\mu_{\ob{n}}}}{\bras{\lambda_1}{\lambda_2}^{n-3}\bras{\lambda_2}{\mu_{\ob{3}}}\bras{\mu_{\ob{4}}}{\mu_{\ob{5}}}...\bras{\mu_{\ob{n-1}}}{\mu_{\ob{n}}}\bras{\mu_{\ob{n}}}{\lambda_1}}
\end{equation}
with fixed spinors $\mu_{\ob{3}\dots \ob{n},a}$ allows us to
reproduce the $n$-point MHV amplitude for gluon ordering $12\ob{3} ... \ob{n}$:
\begin{equation}\label{eq:n-pt}
\mathcal{Q}^{(2,n-2)}_{\mu,\ot{\mu}}\left[\eta_{\mu}^{(\mathbf{1})} \,\eta^{(12,\ob{3}\dots\ob{n})}\right]\propto \mathcal{A}^{(12,3\dots n)}_{\mu,\ot{\mu}}.
\end{equation}

Analogously, it is possible  to reproduce  the charge
$\mathcal{Q}^{(2,n-2)}_{\mu,\ot{\mu}}\left[\eta_{\mu}^{(\sigma)} \,
\eta^{(12,\ob{3}\dots\ob{n})}\right]$ with any other ordering $\sigma(1)...\sigma(n)$
parameterized by $\sigma\in\mathfrak{S}_n$ by choosing parameter
\begin{equation}\label{eq:toMHV_n}
\eta_\mu^{(\sigma)} = \left.\dfrac{\bras{\lambda_1}{\mu_{\ob{3}}}...\bras{\lambda_1}{\mu_{\ob{n}}}\cdot \bras{\lambda_2}{\mu_{\ob{3}}}...\bras{\lambda_2}{\mu_{\ob{n}}}}{\bras{\lambda_1}{\lambda_2}^{n-4}\bras{\nu_{\sigma{(1)}}}{\nu_{\sigma{(2)}}}\bras{\nu_{\sigma{(2)}}}{\nu_{\sigma{(3)}}}...\bras{\nu_{\sigma{(n)}}}{\nu_{\sigma{(1)}}}}\right|_{\nu_{1,2}\to\lambda_{1,2},\, \nu_{3\dots n}\to \mu_{\ob{3}\dots \ob{n}}}\,.
\end{equation}
Note that except for the cases $n=3$ or $4$ for the ordering $1\ob{3}2\ob{4}$, expressions \eqref{eq:toMHV_n1} and \eqref{eq:toMHV_n} contain momentum variables $\lambda_{1,2}$ in the denominator and therefore  the MHV $n$-point amplitude
is reproduced by
$\mathcal{Q}^{(2,n-2)}_{\mu,\ot{\mu}}\left[\eta^{(\sigma)}_\mu\eta^{(12,\ob{3}\dots\ob{n})}\right]$ with non-local parameter $\eta$.

\subsection{MHV 3-point amplitude for higher spins}

Let us also consider MHV-like scattering  for three massless HS fields with helicities $ - s_1 <0$, $- s_2 <0$ and $s_3 >0$ such that $s_1 + s_2 - s_3 > 0$ \cite{BC_hs,Hamed_hs}. 
Note that the latter condition implies that spins of particles obey the inequality $s_3 < s_1 + s_2$ known to play significant role in the context of locality of HS interactions since \cite{GY} (see  also \cite{locality,homotopy} and references therein). Such amplitude has the form\footnote{Construction of HS interaction vertices within the light-cone approach was initiated in \cite{BBB}. The problem of classification of all HS cubic vertices in flat $3+1$-dimensional flat space-time was solved in \cite{BBL} and generalized to any $D$ in \cite{FraMets}.}  \cite{BC_hs,Hamed_hs}
\begin{equation}\label{eq:PTf_3hs}
\mathcal{A}^{(1_{s_1}\,2_{s_2},3_{s_3})}_{\mu,\ot{\mu}} = \langle{\mu_1\mu_2\rangle}^{(s_1 + s_2 + s_3)}\langle\mu_2\mu_3\rangle^{(s_2 - s_1 - s_3)} \langle\mu_3\mu_1\rangle^{(s_1 - s_2 - s_3)}\,\delta^{(4)}\big(\sum_{I=1}^3 \mu_{Ia} \ot{\mu}_{I\p{a}}\big).
\end{equation}
To reproduce this amplitude we take \eqref{eq:amplitude_3pt} with the modified parameter
\begin{equation}\label{eq:parameter_hs}
\eta^{(1_{s_1}\,2_{s_2},\ob{3}_{s_3})} = \langle\lambda_1 \,\lambda_2\rangle^{\sum_{I=1}^3(s_I - 1)}\,\rho_{1\ob{3}}^{(s_1 - 1) + (s_3 - 1) - (s_2 - 1)}\,\rho_{2\ob{3}}^{(s_2 - 1) + (s_3 - 1) - (s_1 - 1)}\,\sgn{\rho_{1\ob{3}}}\,\sgn{\rho_{2\ob{3}}}
\end{equation}
which reduces to $\eta^{(12,\ob{3})}$ in \eqref{eq:amplitude_3pt} for $s_{1,2,3} = 1$. To perform integration \eqref{eq:amplitude}
\begin{multline}\label{eq:amplitude_hs}
\mathcal{Q}^{(1_{s_1}\,2_{s_2},\ob{3}_{s_3})}_{\mu,\ot{\mu}}\left[\eta^{(1_{s_1}\,2_{s_2},\ob{3}_{s_3})}\right] = \int \D^4\lambda\,\D^{2}\ot{\lambda}\,\D^{2}y\,\D^{4}\ot{y}\;\mathrm{exp}\left(i\ot{\mu}_{\ind{i}}\ot{y}_{\ind{i}} + i\mu_{\ob{3}}y_{\ob{3}}\right)\\ \eta^{(1_{s_1}\,2_{s_2},\ob{3}_{s_3})}\,\delta^4\big(\lambda-\mu\big)\,\delta^{2}\big(\ot{\lambda}-\ot{\mu}\big)
\end{multline}
we make use of the following consequence of \eqref{eq:formula} \cite{Gelf_Shil}:
\begin{equation}
\int_{-\infty}^{+\infty} x^n\,\sgn{x}\,e^{ikx}\D x = \dfrac{2i^{n+1}\,n!}{k^{n+1}},\;n=0,1,2,\dots .
\end{equation}
This establishes equivalence of \eqref{eq:amplitude_hs} and \eqref{eq:PTf_3hs}
\begin{equation}
\mathcal{Q}^{(1_{s_1} 2_{s_2},\ob{3}_{s_3})}_{\mu,\ot{\mu}}\left[\eta^{(1_{s_1}\,2_{s_2},\ob{3}_{s_3})}\right] \propto \mathcal{A}^{(1_{s_1} 2_{s_2},3_{s_3})}_{\mu,\ot{\mu}}.
\end{equation}

Note that spins of particles are controlled by the powers of variables $\rho_{\ind{i}\oind{j}}$ in the parameter $\eta$ \eqref{eq:parameter_hs}
 in agreement with the  fact that helicity in HS theory  is associated with the 
difference of degrees of homogeneity of the generating field $\fld{1}{y,\ot{y}}{x}$ in  $y$ and $\ot{y}$  \cite{vasiliev_starprod_ads}.

\subsection{$x$-space integration}

It is also instructive to perform integration of \eqref{eq:form_helicity} over a surface $\mathbf{S}_2$ containing space-time to show that the result coincides with that for integration over $\mathbf{S}_1$ at $x = 0$. To  this end, for  $n$-particle amplitude (Section \ref{n-pt}), one has to choose $2(n-2)$ variables in addition to the $4$ variables $x^{a\p{a}}$  to realize a pullback from $2n$ variables $y^{a}_{\oind{j}} + x^{a\p{a}}\ot{\lambda}_{\oind{j}\p{a}}$, $\ot{y}^{\p{a}}_{\ind{i}} + x^{a\p{a}}\lambda_{\ind{i}a}$ (\textit{cf.} \eqref{eq:characteristics} and \eqref{eq:form}). This can be achieved by considering equations \eqref{eq:link}   expressing at given $\lambda_{\ind{i}a}$ and $\ot{\lambda}_{\oind{i}\p{a}}$ $2n$ variables $y,\ot{y}$ in terms of $2(n-2)$ variables $\rho_{\ind{i}\oind{j}}$ and $4$ variables $\ot{\xi}^{\p{a}}_{\ind{i}}$ parametrizing solutions to the homogeneous equations at $\rho_{\ind{i}\oind{j}} = 0$ (\textit{cf.} \eqref{eq:reparametrization}). This change of variables is non-degenerate for all $\lambda_{\ind{i}a},\ot{\lambda}_{\oind{i}\p{a}}$ except for the measure-zero set  $\det \lambda_{\ind{i} a} = 0$.

The shift
\begin{equation}\label{eq:char_shift}
y^{a}_{\oind{i}}\to y^{a}_{\oind{i}} + x^{a\p{a}}\ot{\lambda}_{\oind{i}\p{a}},\quad \ot{y}^{\p{a}}_{\ind{i}}\to \ot{y}^{\p{a}}_{\ind{i}} + x^{a\p{a}}\lambda_{\ind{i}\p{a}}
\end{equation}
leaves $\rho_{\ind{i}\oind{j}}$ invariant \eqref{eq:link}. As a result, integration over $x$-space is equivalent to the integration over variables $\ot{\xi}$ (\textit{cf.} \eqref{eq:reparametrization} and \eqref{eq:jacobian_n}) with the following reparametrization
\begin{equation}
\ot{\xi}^{\p{a}}_{\ind{i}} = x^{a\p{a}} \lambda_{\ind{i}a}\;\Rightarrow\; \D^4\ot{\xi} = \langle \lambda_1\lambda_2\rangle^2 \,\D^4 x
\end{equation}
bringing \eqref{eq:amplitude_n} to the form
\begin{multline}\label{eq:amplitude_nx}
\mathcal{Q}^{(2,n-2)}_{\mu,\ot{\mu}}\left[\eta^{(12,\ob{3}\dots\ob{n})}\right] =
 \int \D^4\lambda\,\D^{2(n-2)}\ot{\lambda}\,\D^{2(n-2)}\rho\;\D^{4}x\;
 \dfrac{1}{\langle\lambda_1\lambda_2\rangle^{n-4}}\,\exp
 \left(i\sum_{I=1}^{n}\mu_{Ia}\ot{\mu}_{I\p{a}}x^{a\p{a}}\right)\\ \exp\left(i\,\rho_{\ind{i}\oind{j}} \,\dfrac{\langle\mu_{\oind{j}} \,\varepsilon_{\ind{i}\ind{k}}\mu_{\ind{k}}\rangle}{\langle \mu_1 \mu_2\rangle}\right)\, \delta^4\big(\lambda-\mu\big)\,\delta^{2(n-2)}\big(\ot{\lambda}-\ot{\mu}\big)\,\prod_{\ind{i},\oind{j}}\sgn\left(\rho_{\ind{i}\oind{j}}\right).
\end{multline}
Integration over $x$-space gives momentum-conservation delta-function, while that over  $\rho$ leads to the result \eqref{eq:MHVn_pre} along the  lines of Section \ref{n-pt}. We conclude that integration
of \eqref{eq:form_helicity} over $\mathbf{S}_1\subset\mathbf{Cor}_2^{(2,n-2)}$, parametrized only by spinor variables, and $\mathbf{S}_2\subset\mathbf{Cor}_2^{(2,n-2)}$ containing space-time variables $x$, lead to the same result though starting from different representations for $\mathcal{Q}^{(2,n-2)}_{\mu,\ot{\mu}}$  \eqref{eq:MHV3_raw2} and \eqref{eq:amplitude_nx}:
\begin{equation}\label{eq:equal}
\left.\mathcal{Q}^{(2,n-2)}_{\mu,\ot{\mu}}\right|_{\mathbf{S}_1} = \left.\mathcal{Q}^{(2,n-2)}_{\mu,\ot{\mu}}\right|_{\mathbf{S}_2}.
\end{equation}
Even in this simplest case the relation between different representations in not obvious unless $\mathbf{S}_1$ and $\mathbf{S}_2$ are regarded as surfaces in the same total space $\mathbf{Cor}_2^{(2,n-2)}$ and each particular representation is obtained via pullback of the on-shell closed differential form \eqref{eq:form}. That differential form \eqref{eq:form} is closed ensures that the result of integration is insensitive to the change of integration surface within the same homotopy class.

 Note that the condition $\rk \ge 3$ is essential  for representation \eqref{eq:amplitude_nx} and  expression \eqref{eq:amplitude_n} to make sense. This is because $4d$ momentum-conservation $\delta$-function  is ill-defined for $\rk = 2$, \textit{i.e.}, for the conservation law $\delta^{(4)}\left(p_1 + p_2\right)$ with both $p_1$ and $p_2$ being on-shell. This conforms to the fact  \cite{theta} that the usual bilinear HS conserved charges with $\rk = 2$ are represented as space-time integrals over a $3$-dimensional space-like surface with the fourth dimension being compact manifold carrying spin. For higher ranks $\rk \ge 3$ it is possible to perform integration over the four-dimensional $x$-space to obtain a well-defined expression $\delta^{(4)}\left(\sum_{I=1}^{\rk}p_I\right)$ for conservation law of the sum of $\rk$ on-shell momenta (\textit{cf.} \eqref{eq:amplitude_nx}).

 More in detail, consider the case $\rk = 2$ with $m = \ob{m} = 1$ with
 characteristics (\ref{eq:characteristics})
\begin{equation}\label{eq:characteristics_r2}
    y^a + x^{a\p{a}}\ot{\lambda}_{\p{a}}\quad\text{and}\quad \ot{y}^{\p{a}} + x^{a\p{a}}\lambda_{a},
\end{equation}
the ``link variable'' $\rho$ \eqref{eq:link}
\begin{equation}\label{eq:link_r2}
    \rho = \lambda_{a} y^a - \ot{\lambda}_{\p{a}}\ot{y}^{\p{a}}
\end{equation}
and conserved charge \eqref{eq:amplitude} with $\eta^{(1,\ob{1})} = \sgn{\rho}$
\begin{multline}\label{eq:amplitude_charge}
\mathcal{Q}^{(1,1)}_{\mu,\ot{\mu}}\left[\eta^{(1,\ob{1})}\right] = \int_{\mathbf{S}} \D^2\lambda\,\D^2\ot{\lambda}\, \D^2\big(y + x\ot{\lambda}\big)\,\D^2\big(\ot{y} + x\lambda\big)\, \exp\left[i\left(\mu_{\ob{1}a}y^{a} + \ot{\mu}_{1\p{a}}\ot{y}^{\p{a}}\right)\right]\\ \prod_{a=1,2}\delta\big(\lambda_{1a} - \mu_{1a}\big)\,\prod_{\p{a}=1,2}\delta\big(\ot{\lambda}_{\ob{1}\p{a}} - \ot{\mu}_{\ob{1}\p{a}}\big)\,\sgn{\big(\lambda_a y^a - \ot{\lambda}_{\p{a}}\ot{y}^{\p{a}}\big)}.
\end{multline}
The easiest way to compute \eqref{eq:amplitude_charge} is to use the integration surface $\mathbf{S}_1\subset \mathbf{Cor}^{(1,1)}_2$ at $x = 0$. This yields
\begin{multline}\label{eq:amplitude_charge_ans}
    \mathcal{Q}^{(1,1)}_{\mu,\ot{\mu}}\left[\eta^{(1,\ob{1})}\right] \propto \left(\dfrac{\mu_{11}}{\mu_{\ob{1}1}}\right)^2\,\big(\mu_{11}\ot{\mu}_{1\p{1}} - \mu_{\ob{1}2}\ot{\mu}_{\ob{1}\p{2}}\big)\cdot \\ \delta\big(\mu_{11}\ot{\mu}_{1\p{2}} + \mu_{\ob{1}1}\ot{\mu}_{\ob{1}\p{2}}\big)\,\delta\big(\mu_{12}\ot{\mu}_{1\p{1}} + \mu_{\ob{1}2}\ot{\mu}_{\ob{1}\p{1}}\big)\,\delta\big(\mu_{11}\ot{\mu}_{1\p{1}} - \mu_{12}\ot{\mu}_{1\p{2}} + \mu_{\ob{1}1}\ot{\mu}_{\ob{1}\p{1}} - \mu_{\ob{1}2}\ot{\mu}_{\ob{1}\p{2}} \big).
\end{multline}
Delta functions on the {\it rhs} of \eqref{eq:amplitude_charge_ans} express conservation condition for the $3$-dimensional space-like projection\footnote{Recall that vector-spinor dictionary is established by $2\times 2$ Hermitian matrices $\sigma_{\nu\,a\p{a}}$ ($\nu = 0\dots 3$) with $P_{a\p{a}} = P^{\nu}\,\sigma_{\nu\,a\p{a}}$.} of the total momentum $P_{a\p{a}} = \mu_{1a}\ot{\mu}_{1\p{a}} + \mu_{\ob{1}a}\ot{\mu}_{\ob{1}\p{a}}$. Hence, at  $\rk = 2$ amplitude-like expressions for the conserved charges \eqref{eq:amplitude} contain $3$-dimensional delta-function as in \eqref{eq:amplitude_charge_ans}. As mentioned in Section \ref{Sec_HS}, $\rk=2$-systems describe bilinear conserved HS charges and therefore expression \eqref{eq:amplitude_charge_ans} describes free propagation of a particle.

The three-dimensional delta-function in \eqref{eq:amplitude_charge_ans} can be obtained as a $3d$ space integral as follows. After integration over $\lambda,\ot{\lambda}$ in \eqref{eq:amplitude_charge} one is left with the integration of a top-form in characteristics \eqref{eq:characteristics_r2}
\begin{equation}\label{eq:characteristics_r2_topform}
    \D^2 \big(y + x\ot{\mu}_{\ob{1}}\big)\,\D^2 \big(\ot{y} + x\mu_1\big).
\end{equation}
It is straightforward to verify that the pullback of \eqref{eq:characteristics_r2_topform} to $\Mink{2}$ is zero, while that to any surface $\mathbf{S}_2\subset \mathbf{Cor}^{(1,1)}_2$ containing a $3d$ surface in $\Mink{2}$ is
\begin{equation}\label{eq:characteristics_r2_pullback}
   \left. \D^2 \big(y + x\ot{\mu}_{\ob{1}}\big)\,\D^2 \big(\ot{y} + x\mu_1\big)\right|_{\mathbf{S}_2} \propto \D \rho\;\D^3x^{a\p{a}}\mu_{1a}\ot{\mu}_{\ob{1}\p{a}}\,,\qquad
   \D^3x^{a\p{a}} := \D x^{a}{}_{\p{c}}\,\D x_{c}{}^{\p{c}}\,\D x^{c\p{a}}\,.
\end{equation}
By taking space-like surface in $\Mink{2}$ subject to condition $x^{1\p{1}} + x^{2\p{2}} = 0$ and parametrization of $\rho = \mu_{11}y^{1}$ \eqref{eq:link_r2} one arrives at \eqref{eq:amplitude_charge} in the form ({\it cf.} \eqref{eq:formula})
\begin{multline}\label{eq:amplitude_charge_ans2}
    \mathcal{Q}^{(1,1)}_{\mu,\ot{\mu}}\left[\eta^{(1,\ob{1})}\right] \propto \int_{\mathbf{S}_2} \D\rho\,\D^3 x^{a\p{a}}\,\mu_{1a}\ot{\mu}_{\ob{1}\p{a}}\, e^{i\frac{\mu_{\ob{1}1}}{\mu_{11}}\rho}e^{i\,x^{a\p{a}}\left(\mu_{1a}\ot{\mu}_{1\p{a}} + \mu_{\ob{1}a}\ot{\mu}_{\ob{1}\p{a}}\right)}\,\sgn{\rho} \propto\\ \frac{\mu_{11}}{\mu_{\ob{1}1}}\left(\mu_{11}\ot{\mu}_{\ob{1}\p{1}} + \mu_{12}\ot{\mu}_{\ob{1}\p{2}}\right)\, \delta\big(\mu_{11}\ot{\mu}_{1\p{2}} + \mu_{\ob{1}1}\ot{\mu}_{\ob{1}\p{2}}\big)\,\delta\big(\mu_{12}\ot{\mu}_{1\p{1}} + \mu_{\ob{1}2}\ot{\mu}_{\ob{1}\p{1}}\big)\,\delta\big(\mu_{11}\ot{\mu}_{1\p{1}} - \mu_{12}\ot{\mu}_{1\p{2}} + \mu_{\ob{1}1}\ot{\mu}_{\ob{1}\p{1}} - \mu_{\ob{1}2}\ot{\mu}_{\ob{1}\p{2}} \big).
\end{multline}
Expressions \eqref{eq:amplitude_charge_ans} and \eqref{eq:amplitude_charge_ans2} are equal under the $3$-momentum conservation condition.

The above consideration manifests the difference for $x$-integration between the usual bilinear conserved HS charges ($\rk = 2$) and  multi-particle HS charges ($\rk\ge 3$)   most interesting from the perspective of scattering amplitudes.

\subsection{Form-factors for field strengths}\label{Sec_FormFactors}

 In this section we explain why scattering amplitudes constructed in terms of gauge potentials \eqref{eq:gauge_potentials} are reproduced in terms of conserved charges \eqref{eq:amplitude} originally derived in terms of  gauge-invariant field strengths \eqref{eq:unfolded_field}. Let us first spell out the main idea. By  equations \eqref{eq:comst_s}, spin-$s$ field strength is constructed from the $\lfloor s \rfloor$-th derivatives of the spin-$s$ gauge field. This means that one can pass from field strengths to gauge potentials by a non-local transform. HS framework in question contains space-time non-localities in terms of rational dependence of parameters $\eta$ \eqref{eq:eta_uniform} on the variables $\lambda_{\ind{i}a}$ and $\ot{\lambda}_{\oind{j}\p{a}}$ via  \eqref{eq:unfolded_Fourier}. As a result, HS charges corresponding to gauge potentials \eqref{eq:gauge_potentials} are related to those for field strengths \eqref{eq:unfolded_field} by inserting into \eqref{eq:amplitude} an additional parameter \eqref{eq:eta_uniform} containing negative powers of $\lambda_{\ind{i}a},\ot{\lambda}_{\oind{j}\p{a}}$. Though technical details may look bulky because of too many indices we present them for completeness.

Since equation \eqref{eq:unfolded} describes gauge-invariant field strengths (\textit{i.e.,} HS curvatures)
\cite{Vasiliev:1988sa}, conserved charges \eqref{eq:amplitude} most naturally represent $\rk$-particle correlation functions (form-factors) of normally-ordered field operators as
\begin{equation}\label{eq:correlator}
\langle 0\vert :\widehat{C}_{a_1(n_1),\p{a}_1(n^\prime_1)}\left(p_1\right)\dots \widehat{C}_{a_\rk(n_\rk),\dot{a}_\rk(n^\prime_\rk)}\left(p_{\rk}\right):\,\widehat{S}\vert 0\rangle.
\end{equation}
Here $\widehat{S}$ is $S$-matrix\footnote{ The Coleman-Mandula argument \cite{Coleman_Mandula} on the triviality of $S$-matrix with unbroken higher symmetries is not  applicable here because, generally, HS symmetries are broken by a chosen 
 symmetry parameter $\eta$.}, $\widehat{C}_{a(n),\p{a}(n^\prime)}\left(p\right)$ are symmetric spin-tensors in $n$ indices $a$ and  $n^\prime$ indices $\p{a}$, representing the on-shell derivatives of (anti-)self-dual components of the generalized spin-$s$ Weyl tensors  in momentum space. (Anti-)self-duality is determined by $\mathrm{sign}\left(n^\prime - n\right) = (+)-$. As explained in Section \ref{Sec_1_1}, generalized Weyl tensors are related to gauge fields \eqref{eq:gauge_potentials} via  Central On-Mass-Shell Theorem \eqref{eq:comst}. More in detail, let gauge potentials \eqref{eq:gauge_potentials} be of the plane-wave form with the following (negative) polarization spin-tensors:
\begin{equation}\label{eq:wave_potentials}
\D x^{a\p{a}}\,w_{a(k),\p{a}(l)}^{(-)}\left(x\right) \propto \theta\left(k-l\right)\,\D x^{a\p{a}}\,\dfrac{\overbrace{\mu_{a}\dots\mu_{a}}^{k}\,\overbrace{\ot{\varkappa}_{\p{a}}\dots \ot{\varkappa}_{\p{a}}}^{l}}{\braq{\ot{\varkappa}}{\ot{\mu}}^{l}}\,e^{i\,\mu_{b}\ot{\mu}_{\p{b}}x^{b\p{b}}},
\end{equation}
where $\mu_{a}\ot{\mu}_{\p{a}}$ is light-like momentum and $\ot{\varkappa}_{\p{a}}$ is an arbitrary reference spinor such that $\braq{\ot{\varkappa}}{\ot{\mu}} \neq 0$ ({\it cf.} Section \ref{Sec_1_1}). Then fields \eqref{eq:wave_potentials} obey equations \eqref{eq:comst_1}, \eqref{eq:comst_s} with the generalized Weyl tensors
\begin{equation}\label{eq:C}
c_{a(n),\p{a}(n^\prime)}^{(-)}\left(x\right) \propto \theta\left(n-n^{\prime}\right)\underbrace{\mu_a\dots\mu_a}_{n}\,
\underbrace{\ot{\mu}_{\p{a}}\dots\ot{\mu}_{\p{a}}}_{n^\prime}\,e^{i\,\mu_{b}
\ot{\mu}_{\p{b}}x^{b\p{b}}}\,,
\end{equation}
where $n-n^\prime = 2s$. As a result, gauge potentials \eqref{eq:wave_potentials} correspond to the self-dual (SD) projection of free HS fields. Construction of (positive-polarization) anti-self-dual (ASD) counterparts $W^{(+)}$ for \eqref{eq:wave_potentials} and $C^{(+)}$ for \eqref{eq:C} is analogous.

Upon quantization negative polarization spin-tensors of $W^{(-)}$ \eqref{eq:wave_potentials} and $C^{(-)}$ \eqref{eq:C}, as well as $W^{(+)}$ and $C^{(+)}$, are equipped with creation and annihilation operators $\op{a}^{\dagger}_{\pm s} \left(\vec{p}\right),\op{a}_{\pm s}\left(\vec{p}\right)$ of particular momentum, spin and polarization. Right action of the normal ordered products of field operators produce an \textit{out}-state dressed with the polarization tensors as follows
\begin{multline}
\left\langle 0\right| :\op{w}^{(\sigma_1)}_{a_1(k_1),\p{a}_1(l_1)}\left(p_1\right)\dots \op{w}^{(\sigma_\rk)}_{a_\rk(k_\rk),\p{a}_\rk(l_\rk)}\left(p_\rk\right):\; \propto \\ \prod_{I=1}^{\rk}\theta\left(\sgn\left(\sigma\right)\left(l_I - k_I\right)\right)\dfrac{\overbrace{\xi^{(\sigma_I)}_{I\,a_I}\dots \xi^{(\sigma_I)}_{I\,a_I}}^{k_I}\,\overbrace{\ot{\xi}^{(\sigma_I)}_{I\,\p{a}_I}\dots \ot{\xi}^{(\sigma_I)}_{I\,\p{a}_I}}^{l_I}}{\big(\xi^{(\sigma_I)}_I,\ot{\xi}^{(\sigma_I)}_I\big)^{\text{min}\left(k_I,l_I\right)}} \left\langle p_1^{(\sigma_1)}\dots p_\rk^{(\sigma_\rk)}\right|,
\end{multline}
where $\xi^{(\sigma)}_{a} = \theta\left(\sigma\right)\,\varkappa_{a} + \theta\left(-\sigma\right)\,\mu_{a}$ and $\ot{\xi}^{(\sigma)}_{\p{a}} = \theta\left(\sigma\right)\,\ot{\mu}_{\p{a}} + \theta\left(-\sigma\right)\,\ot{\varkappa}_{\p{a}}$, for $p_{a\p{a}} = \mu_{a}\ot{\mu}_{\p{a}}$ and reference spinors $\varkappa_{a}$, $\ot{\varkappa}_{\p{a}}$, with bracket $\big(\xi^{(\sigma)}\,\ot{\xi}^{(\sigma)}\big) := \theta\left(\sigma\right)\bras{\varkappa}{\mu} + \theta\left(-\sigma\right)\braq{\ot{\varkappa}}{\ot{\mu}}$ (\textit{cf.} \eqref{eq:wave_potentials}). Analogous expression takes place for Weyl tensors \eqref{eq:C}:
\begin{multline}
\left\langle 0\right| :\op{C}^{(\sigma_1)}_{a_1(k_1),\p{a}_1(l_1)}\left(p_1\right)\dots \op{C}^{(\sigma_\rk)}_{a_\rk(k_\rk),\p{a}_\rk(l_\rk)}\left(p_\rk\right):\; \propto \\ \prod_{I=1}^{\rk}\underbrace{\mu_{I\,a_I}\dots \mu_{I\,a_I}}_{k_I}\,\underbrace{\ot{\mu}_{I\,\p{a}_I}\dots \ot{\mu}_{I\p{a}_I}}_{l_I} \;\left\langle p_1^{(\sigma_1)}\dots p_\rk^{(\sigma_\rk)}\right|
\end{multline}
with $\sgn{\left(\sigma_I\right)} = \sgn{\left(l_I-k_I\right)}$ (\textit{cf.} \eqref{eq:C}). As a result, form-factors \eqref{eq:correlator} are related to scattering amplitudes via dressing with polarization tensors:
\begin{multline}\label{eq:correlator_amplitude}
\left\langle 0\right| :\op{C}^{(\sigma_1)}_{a_1(k_1),\p{a}_1(l_1)}\left(p_1\right)\dots \op{C}^{(\sigma_\rk)}_{a_\rk(k_\rk),\p{a}_\rk(l_\rk)}\left(p_\rk\right):\,\op{S}\left|0\right\rangle\; \propto \\ \prod_{I=1}^{\rk}\underbrace{\mu_{I\,a_I}\dots \mu_{I\,a_I}}_{k_I}\,\underbrace{\ot{\mu}_{I\,\p{a}_I}\dots \ot{\mu}_{I\p{a}_I}}_{l_I}\;\big\langle p_1^{(\sigma_1)}\dots p_\rk^{(\sigma_\rk)}\big|\op{S}\big|0\big\rangle\,.
\end{multline}
For instance, consider correlators for spin-$1$ field strengths. At the tree level, form-factors containing field strengths are related to the $\rk$-particle Parke-Taylor amplitude \eqref{eq:PTf}
\begin{multline}\label{eq:correlator_MHV}
\left\langle 0\right| :\op{C}^{(-)}_{a_1 a_1}\left(p_1\right)\,\op{C}^{(-)}_{a_2 a_2}\left(p_2\right)\,\op{C}^{(+)}_{\p{a}_{\ob{3}} \p{a}_{\ob{3}}}\left(p_3\right)\dots \op{C}^{(+)}_{\p{a}_{\ob{\rk}} \p{a}_{\ob{\rk}}}\left(p_\rk\right):\,\op{S}\left|0\right\rangle_{\text{tree}} \propto \\ \prod_{\ind{i}=1,2}\mu_{\ind{i},a_{\ind{i}}}\mu_{\ind{i},a_{\ind{i}}}\prod_{\oind{j}=\ob{3}\dots\ob{\rk}} \ot{\mu}_{\oind{j},\p{a}_{\oind{j}}}\ot{\mu}_{\oind{j},\p{a}_{\oind{j}}}\;\dfrac{\bras{\mu_1}{\mu_2}^4}{\bras{\mu_1}{\mu_2}\bras{\mu_2}{\mu_{\ob{3}}}\dots \bras{\mu_{\ob{\rk}}}{\mu_1}} \delta^{(4)}\big(\sum_{I=1,2,\ob{3}\dots\ob{\rk}} \mu_{Ia}\ot{\mu}_{I\dot{a}}\big).
\end{multline}
Therefore correlators \eqref{eq:correlator_MHV} can be obtained as conserved charges \eqref{eq:n-pt} by introducing additional parameter
 $\eta^{C}_{a_1(2),a_2(2),\p{a}_{\ob{3}}(2),\dots, \p{a}_{\ob{\rk}}(2)} =
 \prod_{\ind{i}=1,2}\lambda_{\ind{i},a_{\ind{i}}}
 \lambda_{\ind{i},a_{\ind{i}}}\prod_{\oind{j}=\ob{3}\dots\ob{\rk}}
 \ot{\lambda}_{\oind{j},\p{a}_{\oind{j}}}\ot{\lambda}_{\oind{j},\p{a}_{\oind{j}}}$ to \eqref{eq:n-pt},
\begin{multline}\label{eq:formfactor_c}
\mathcal{Q}^{(2,\rk-2)}_{\mu,\ot{\mu}}\left[\eta^{C}_{a_1(2)a_2(2)\p{a}_{\ob{3}}(2)\dots \p{a}_{\ob{\rk}}(2)} \,\eta^{(\mathbf{1})}\,\eta^{(12,\ob{3}\dots\ob{\rk})}\right] \propto \\ \left\langle 0\right| :\op{C}^{(-)}_{a_1 a_1}\left(p_1\right)\,\op{C}^{(-)}_{a_2 a_2}\left(p_2\right)\,\op{C}^{(+)}_{\p{a}_{\ob{3}} \p{a}_{\ob{3}}}\left(p_3\right)\dots \op{C}^{(+)}_{\p{a}_{\ob{\rk}} \p{a}_{\ob{\rk}}}\left(p_\rk\right):\,\op{S}\left|0\right\rangle_{\text{tree}}.
\end{multline}
Analogous expression for the electromagnetic potential (\textit{cf.} \eqref{eq:wave_potentials} at $k = l = 1$) reads as
\begin{multline}\label{eq:correlator_MHV_w}
\left\langle 0\right| :\op{w}^{(-)}_{a_1 \p{a}_1}\left(p_1\right)\,\op{w}^{(-)}_{a_2 \p{a}_2}\left(p_2\right)\,\op{w}^{(+)}_{a_{\ob{3}} \p{a}_{\ob{3}}}\left(p_3\right)\dots \op{w}^{(+)}_{a_{\ob{\rk}} \p{a}_{\ob{\rk}}}\left(p_\rk\right):\,\op{S}\left|0\right\rangle_{\text{tree}} \propto \\ \prod_{\ind{i}=1,2}\dfrac{\mu_{\ind{i},a_{\ind{i}}}\ot{\varkappa}_{\ind{i},\p{a}_{\ind{i}}}}{\braq{\ot{\varkappa}_{\ind{i}}}{\ot{\mu}_\ind{i}}}\prod_{\oind{j}=\ob{3}\dots\ob{\rk}} \dfrac{\ot{\mu}_{\oind{j},\p{a}_{\oind{j}}}\varkappa_{\oind{j},a_{\oind{j}}}}{\bras{\varkappa_{\oind{j}}}{\mu_{\oind{j}}}}\;\dfrac{\bras{\mu_1}{\mu_2}^4}{\bras{\mu_1}{\mu_2}\bras{\mu_2}{\mu_{\ob{3}}}\dots \bras{\mu_{\ob{\rk}}}{\mu_1}} \delta^{(4)}\big(\sum_{I=1,2,\ob{3}\dots\ob{\rk}} \mu_{Ia}\ot{\mu}_{I\dot{a}}\big).
\end{multline}
It can be expressed as conserved charge \eqref{eq:n-pt} via insertion of the
additional parameter $\eta^{w}_{a_1\p{a}_1, a_2\p{a}_2, a_{\ob{3}}\p{a}_{\ob{3}},\dots, a_{\ob{\rk}}\p{a}_{\ob{\rk}}} = \prod_{\ind{i}=1,2}\dfrac{\lambda_{\ind{i},a_{\ind{i}}}\ot{\varkappa}_{\ind{i},\p{a}_{\ind{i}}}}{{\braq{\ot{\varkappa}_{\ind{i}}}{\ot{\mu}_\ind{i}}}}\prod_{\oind{j}=\ob{3}\dots\ob{\rk}}\dfrac{\ot{\lambda}_{\oind{j},\p{a}_{\oind{j}}}\varkappa_{\oind{j},a_{\oind{j}}}}{\bras{\varkappa_{\oind{j}}}{\mu_{\oind{j}}}}$:
\begin{multline}\label{eq:formfactor_w}
\mathcal{Q}^{(2,\rk-2)}_{\mu,\ot{\mu}}\left[\eta^{w}_{a_1\p{a}_1, a_2\p{a}_2, a_{\ob{3}}\p{a}_{\ob{3}},\dots, a_{\ob{\rk}}\p{a}_{\ob{\rk}}} \,\eta^{(\mathbf{1})}\,\eta^{(12,\ob{3}\dots\ob{\rk})}\right] \propto \\ \left\langle 0\right| :\op{w}^{(-)}_{a_1 \p{a}_1}\left(p_1\right)\,\op{w}^{(-)}_{a_2 \p{a}_2}\left(p_2\right)\,\op{w}^{(+)}_{a_{\ob{3}} \p{a}_{\ob{3}}}\left(p_3\right)\dots \op{w}^{(+)}_{a_{\ob{\rk}} \p{a}_{\ob{\rk}}}\left(p_\rk\right):\,\op{S}\left|0\right\rangle_{\text{tree}}.
\end{multline}

We conclude that form-factors for both gauge invariant Weyl tensors and gauge potentials can be represented as multi-particle HS conserved charges \eqref{eq:amplitude}. The action of gauge transformations on potentials \eqref{eq:gauge_potentials} turns into the action on symmetry parameter $\eta$ in \eqref{eq:amplitude} according to \eqref{eq:gauge_transform} (for the spin-$1$ case it amounts to shifting reference spinors $\varkappa_{\ind{i}a}\rightarrow \varkappa_{\ind{i}a} + \epsilon \mu_{\ind{i}a}$, $\ot{\varkappa}_{\oind{j}\p{a}}\rightarrow \ot{\varkappa}_{\oind{j}\p{a}} + \epsilon\ot{\mu}_{\oind{j}\p{a}}$).

Note that charges \eqref{eq:formfactor_w} for Weyl tensors are related to those for gauge potentials \eqref{eq:formfactor_c} via a non-local  transformation. Namely, for a given momentum $p_{a\p{a}} = \mu_a\ot{\mu}_{\p{a}}$ with reference spinors $\varkappa_a$, $\ot{\varkappa}_{\p{a}}$ consider parameters
\begin{equation}\label{eq:parameters_nonloc}
\eta^{(-)}_{a\p{a}}\big(\lambda\big) = \dfrac{\varkappa_a\ot{\varkappa}_{\p{a}}}{\bras{\lambda}{\varkappa}\braq{\ot{\varkappa}}{\ot{\mu}}}\quad \text{and} \quad\eta^{(+)}_{a\p{a}}\big(\ot{\lambda}\big) = \dfrac{\varkappa_a\ot{\varkappa}_{\p{a}}}{\bras{\mu}{\varkappa}\braq{\ot{\varkappa}}{\ot{\lambda}}}.
\end{equation}
Then it is straightforward to see that
\begin{multline}\label{eq:relation_cw}
\mathcal{Q}^{(2,\rk-2)}_{\mu,\ot{\mu}}\left[\eta^{w}_{a_1\p{a}_1, a_2\p{a}_2, a_{\ob{3}}\p{a}_{\ob{3}},\dots, a_{\ob{\rk}}\p{a}_{\ob{\rk}}} \,\eta^{(\mathbf{1})}\,\eta^{(12,\ob{3}\dots\ob{\rk})}\right] \propto \\ \mathcal{Q}^{(2,\rk-2)}_{\mu,\ot{\mu}}\left[\eta^{(-)}_{c_1\p{a}_1}\big(\lambda_1\big)\eta^{(-)}_{c_2\p{a}_2}\big(\lambda_2\big)\eta^{(+)}_{a_{\ob{3}}\p{c}_{\ob{3}}}\big(\ot{\lambda}_{\ob{3}}\big)\dots \eta^{(+)}_{a_{\ob{\rk}}\p{c}_{\ob{\rk}}}\big(\ot{\lambda}_{\ob{\rk}}\big)\,\eta^{C}{}_{a_1}{}^{c_1}{}_{,a_2}{}^{c_2}{}_{,\p{a}_{\ob{3}}}{}^{\p{c}_{\ob{3}}}{}_{,\dots , a_{\ob{\rk}}}{}^{\p{c}_{\ob{\rk}}} \,\eta^{(\mathbf{1})}\,\eta^{(12,\ob{3}\dots\ob{\rk})}\right].
\end{multline}
This relation justifies application of the multi-particle conserved HS charges \eqref{eq:amplitude} constructed originally in terms  of gauge-invariant zero-forms \eqref{eq:unfolded_field} to calculation of scattering  amplitudes  in terms of gauge fields \eqref{eq:gauge_potentials}. Since zero-forms are related to gauge potentials via space-time differentiation, potentials are reconstructed in terms of zero-forms via a non-local (and gauge-dependent) transformation. In accordance with unfolded equation \eqref{eq:unfolded_Fourier} space-time non-locality results from dividing by spinor variables $\lambda_{\ind{i}a}$, $\ot{\lambda}_{\oind{j}\p{a}}$ (cf. \eqref{eq:parameters_nonloc}, \eqref{eq:relation_cw}). Since amplitudes are non-local anyway, such transformations are allowed in this framework.

\section{Generalized Minkowski space}\label{Sec_Mink}

Rank-$\rk$ equation \eqref{eq:unfolded} can be generalized by extending the range of values for spinorial indices
\begin{equation}\label{eq:unfolded_K}
\left(\dfrac{\partial}{\partial x^{\got{a}\pgot{a}}} + i\,\delta_{IJ}\dfrac{\partial^2}{\partial y^{\got{a}}_I\partial \ot{y}^{\pgot{a}}_J}\right)\fld{\rk}{y,\ot{y}}{x} = 0,
\end{equation}
where $I,J=1\dots\rk$ and $\got{a},\pgot{a} = 1\dots K$, $K\ge 2$. Coordinates $x^{\got{a}\pgot{a}}$
are said to parameterize \textit{generalized Minkowski space-time} $\Mink{K}$
\cite{higher_rank}.
Equation \eqref{eq:unfolded_K} is known to admit physical interpretation for some special values of $\rk$ and $K$: for $\rk = 1$, $K = 2,4,8$ it describes HS multiplets in flat space of various dimensions \cite{Mar,Bandos:2005mb}
and for $\rk = 2$, $K = 2$ it describes conformal conserved currents in $d=4$
Minkowski space \cite{Gelfond:2003vh}. For $K=2n$ coordinates $x^{\got{a}\pgot{a}}$ can be represented as a $2n\times 2n$ matrix constituted by $2\times 2$ blocks $x^{a\p{a}}_{\bb{ij}}$, $a,\p{a} = 1,2$ and $\bb{i},\bb{j}=1\dots n$. From this perspective equation
\eqref{eq:unfolded_K} may be regarded as describing dynamics of $n^2$-local fields in $\Mink{2}$ thus giving rise to polylocal conserved charges. Usual Minkowski space $\Mink{2}$ can be regarded as  diagonally embedded subspace of $\Mink{2n}$  with coordinates $x^{a\p{a}}$ of $\Mink{2}$ parametrizing a four-dimensional subspace of $\Mink{2n}$ as $x^{a\p{a}}_{\bb{ij}} = \delta_{\bb{ij}}\,x^{a\p{a}}$.

Conserved charges for polylocal fields \eqref{eq:unfolded_K} of rank $\rk = 1$ with $K = 2n$, $n\ge 3$, can be related to $n$-point MHV amplitudes \eqref{eq:PTf} as follows.  Proceeding along the lines of Section \ref{Sec_HS} equation \eqref{eq:unfolded_K} can be transformed to a first-order PDE (\textit{cf.}
 \eqref{eq:unfolded_Fourier})
\begin{equation}\label{eq:unfolded_Fourier_K}
\left(\dfrac{\partial}{\partial x^{\got{a}\pgot{a}}} -\lambda_{\got{a}}\dfrac{\partial}{\partial \ot{y}^{\pgot{a}}}\right)g^{(1,0)}\big(\lambda,\ot{y}\big| x\big) = 0
\end{equation}
with characteristics
\begin{equation}\label{eq:characteristics_K}
\lambda_{\got{a}},\quad \ot{y}^{\pgot{a}} + x^{\got{a}\pgot{a}}\,\lambda_{\got{a}}.
\end{equation}
Correspondence space $\mathbf{Cor}^{(1,0)}_K = \Mink{K}\times\mathbb{R}^{2K}$ is
parametrized by variables $x^{\got{a}\pgot{a}}$, $\lambda_{\got{a}}$ and
$\ot{y}^{\pgot{a}}$, $\dim\mathbf{Cor}^{(1,0)}_K = K^2 + 2K$. On-shell closed
differential form of degree $2K = 4n$ for solutions $g$ of \eqref{eq:unfolded_Fourier_K} is constructed as follows (\textit{cf.} \eqref{eq:form})
\begin{equation}\label{eq:form_K}
\Omega_K[g] = \D^{2n}\lambda\,\D^{2n}\big(\ot{y} + x\,\lambda\big)\,g\big(\lambda,\ot{y}\big| x\big).
\end{equation}
To bring notations to the form suitable for the diagonal embedding $\Mink{2}\subset\Mink{2n}$
 it is convenient to take characteristics \eqref{eq:characteristics_K} in the
form
\begin{equation}\label{eq:characteristics_K_diag}
\lambda_{\bb{i}a},\quad \ot{y}_{\bb{i}}^{\p{a}} + x^{a\p{a}}_{\bb{ij}}\,\lambda_{\bb{j}a},
\end{equation}
where $a,\p{a} = 1,2$ and $\bb{i},\bb{j} = 1\dots n$ enumerate single-particle sectors in $\Mink{2}$. 

To obtain MHV amplitudes \eqref{eq:PTf} by integration of \eqref{eq:form_K} one has to break $\mathfrak{S}_n$ symmetry acting on single-particle variables \eqref{eq:characteristics_K_diag} by permutations. To this end we split index $\bb{i} = 1\dots n$ into its $2$ values $\ind{i} = 1,2$ and $n-2$ values $\oind{j} = \ob{3}\dots \ob{n},$\footnote{To describe different positions of two negative-helicity gluons, one can choose any $2$ values of $1\dots n$ for the index $\ind{i}$ and the rest $n-2$ ones for $\oind{j}$ } thus explicitly distinguishing between variables $\lambda_{\ind{i}a}$, $\ot{y}_{\ind{i}}^{\p{a}} + x^{a\p{a}}_{\ind{i}\bb{j}} \lambda_{\bb{j}a}$ and $\lambda_{\oind{j}a}$, $\ot{y}_{\oind{j}}^{\p{a}} + x^{a\p{a}}_{\oind{j}\bb{i}}\lambda_{\bb{i}a}$ (summation over repeated indices is implied). To obtain the MHV $n$-point amplitude via pullback of \eqref{eq:form_K} to the diagonal Minkowski subspace $\Mink{2}\subset\Mink{2n}$ it suffices to specify $g^{(1,0)} = g^{(12,\ob{3}\dots\ob{n})}_{\mu,\ot{\mu}}$ parametrized by momentum spinors $\mu_{\bb{i}a}$, $\ot{\mu}_{\bb{i}\p{a}}$ in \eqref{eq:form_K} as follows
\begin{multline}\label{eq:solution_K}
g^{(12,\ob{3}\dots\ob{n})}_{\mu,\ot{\mu}}\big(\lambda,\ot{y}\big| x\big) =
\dfrac{\bras{\lambda_1}{\lambda_2}^4}{\bras{\lambda_1}{\lambda_2}
\bras{\lambda_2}{\lambda_{\ob{3}}}\dots \bras{\lambda_{\ob{n}}}{\lambda_1}}\,
\delta^{(4)}\left(\sigma_{\ind{ij}}\right)\, \delta^{(2n)}
 \big(\lambda_{\bb{i}a} - \mu_{\bb{i}a}\big) \\ \delta^{(4)}\big(\lambda_{\ind{i}a} +
 \sigma_{\ind{i}\oind{j}}\,\lambda_{\oind{j}a}\big)\,\delta^{(2(n-2))}\left(\dfrac{\ot{\mu}_{\oind{j}\p{a}} - \sigma_{\ind{i}\oind{j}}\,\ot{\mu}_{\ind{i}\p{a}}}{\braq{\ot{\mu}_1}{\ot{\mu}_2}}\right),
\end{multline}
where $\sigma_{\ind{i}\bb{j}} = \ot{\mu}_{\ind{i}\p{a}}\big(\ot{y}_{\bb{j}}^{\p{a}} + x^{a\p{a}}_{\bb{j}\bb{k}}\lambda_{\bb{k}a}\big)$. The form of expression \eqref{eq:solution_K} is analogous to the construction of integration over Grassmannian proposed in \cite{Hamed_S}. It manifestly breaks $\mathfrak{S}_n$ symmetry. The easiest way to check that integration of differential form \eqref{eq:form_K} with solution \eqref{eq:solution_K} indeed leads to the
$n$-particle MHV  amplitude \eqref{eq:PTf},
\begin{equation}\label{eq:result_MHV}
\int \Omega_{2n}\big[g^{(12,\ob{3}\dots\ob{n})}_{\mu,\ot{\mu}}\big] \propto \mathcal{A}_{\mu,\ot{\mu}}^{(12,3\dots n)},
\end{equation}
 is by choosing integration surface $\mathbf{S}_1\subset\mathbf{Cor}^{(1,0)}_K$ at $x^{a\p{a}}_{\bb{i}\bb{j}} = 0$. The same result is  obtained via diagonal embedding of Minkowski space $\Mink{2}$ with coordinates $x^{a\p{a}}$ into the generalized Minkowski space $\Mink{2n}$ with $x^{a\p{a}}_{\bb{i}\bb{j}} = x^{a\p{a}}\delta_{\bb{i}\bb{j}}$. Here the integration is  over $\mathbf{S}_2\subset \mathbf{Cor}^{(1,0)}_K$ parametrized by $4$ variables $x^{a\p{a}}$ of $\Mink{2}\subset \Mink{2n}$, $2(n-2)$
 variables $\ot{y}_{\oind{j}}^{\p{a}}$ and $2n$ variables $\lambda_{\bb{i}a}$ with $\ot{y}_{\ind{i}}^{\p{a}} = 0$.

Another interesting choice of integration surface $\mathbf{S}_3\subset \mathbf{Cor}^{(1,0)}_K$ is with $\ot{y}^{\p{a}}_{\bb{i}} = 0$. As for variables $x_{\bb{ij}}^{a\p{a}}$, consider the diagonal embedding of $n$ copies of the usual Minkowski space $\Mink{2}$ into $\Mink{2n}$, $\underbrace{\Mink{2}\times
\dots\times\Mink{2}}_{n}\subset\Mink{2n}$ defined as
\begin{equation}\label{eq:embedding_n}
x_{\bb{ii}}^{a\p{a}} \equiv x_{\bb{i}}^{a\p{a}} \quad\text{and}\quad x_{\bb{ij}}^{a\p{a}} = 0\quad\text{for}\quad \bb{i}\neq\bb{j}.
\end{equation}
In this case one arrives at the pullback of differential form \eqref{eq:form_K}, \eqref{eq:solution_K} to $\underbrace{\Mink{2}\times\dots\times\Mink{2}}_{n}\subset\Mink{2n}$
\begin{equation}\label{eq:pullback_Kx}
\Omega_{2n}[g_{\mu,\ot{\mu}}^{(12,\ob{3}\dots\ob{n})}] \bigg|_{\Mink{2}{}^{\times n}}\propto \D^{2n}\lambda\,\wedge\left(\bigwedge_{\bb{i}=1}^n H_{\bb{i}}^{ab}\lambda_{\bb{i}a}\lambda_{\bb{i}b}\right)\,g^{(12,\ob{3}\dots\ob{n})}_{\mu,\ot{\mu}}\big(\lambda,0\big| x_{\bb{i}}\big),
\end{equation}
where $H_{\bb{i}}^{ab} = \D x_{\bb{i}}^{a}{}_{\p{c}}\wedge\D x_{\bb{i}}^{b\p{c}}$ (no summation over  $\bb{i}$). Note that, in agreement with \cite{Witten}, because of $\delta\left(\lambda_{\bb{i}}-\mu_{\bb{i}}\right)$ in \eqref{eq:solution_K} integration of \eqref{eq:pullback_Kx} over $\lambda$-variables is straightforward leaving one with multiple integration $H_{\bb{i}}^{ab}\,\mu_{\bb{i}a}\mu_{\bb{i}b}$ over $n$ copies of two-dimensional surfaces in $\Mink{2}$ spanned by light-like momenta $p_{\bb{i}\,a\p{a}} = \mu_{\bb{i}a}\ot{\mu}_{\bb{i}\p{a}}$
and negative-helicity polarizations $\epsilon^{(-)}_{\bb{i}\,a\p{a}} = \dfrac{\mu_{\bb{i}a}\ot{\varkappa}_{\bb{i}\p{a}}}{\braq{\ot{\mu}}{\ot{\varkappa}}}$. To perform integration \eqref{eq:result_MHV} it is convenient to use the following parametrization of two-dimensional planes
\begin{equation}\label{eq:plane_SD}
x_{\bb{i}}^{a\p{a}} = \dfrac{\varkappa_{\bb{i}}{}^{a}\ot{z}_{\bb{i}}^{\p{a}}}{\bras{\mu_{\bb{i}}}{\varkappa_{\bb{i}}}},
\end{equation}
where $\varkappa_{\bb{i}a}$ is a constant spinor  for each $\bb{i}$ (no summation over repeated indices $\bb{i}$) and $z_{\bb{i}}^{\p{a}}$ parametrize the $\bb{i}$-th two-dimensional plane in $\Mink{2}$.  The result of integration \eqref{eq:result_MHV} over $\mathbf{S}_3$, as anticipated, coincides with those over $\mathbf{S}_{1}$ and $\mathbf{S}_{2}$. Since representation \eqref{eq:pullback_Kx} contains multiple integration over space-time variables, the resulting charge \eqref{eq:result_MHV} is space-time non-local.
 Note that parametrization \eqref{eq:plane_SD} corresponds to a SD plane in the complexified Minkowski space $\Mink{2}^{\mathbb{C}}$ such that $\D x_{c\p{a}}\left(\ot{z}\right)\wedge\D x^{c}{}_{\p{b}}\left(\ot{z}\right) = 0$ for \eqref{eq:plane_SD}.

\section{Conclusion}\label{Sec_Conclusion}

It is demonstrated that relation between conserved HS charges of tensor powers of free massless fields and scattering amplitudes of QFT conjectured in \cite{higher_rank} indeed takes place. The rationale behind this is that  scattering amplitudes are expressed in terms  of free ingoing and outgoing particles. A general form of the amplitude is determined by an arbitrary function  $\eta$ of characteristics of free unfolded field equations which has the meaning of a HS symmetry parameter \cite{twistors}. Interactions of one or another QFT  determine a theory-dependent form of $\eta$.  In particular, parameter $\eta$ encodes spins of scattering particles \eqref{eq:parameter_hs}.
 That spins of particles are encoded in the HS symmetry parameter via powers of variables $\rho$ \eqref{eq:link} allows one to apply HS framework in question to lower-spin theories as  was explicitly demonstrated for the MHV amplitudes of Yang-Mills theory in  \eqref{eq:result_3pt} and \eqref{eq:n-pt}. A promising novel idea arising from the  relation between multi-particle HS charges and scattering amplitudes is that the infinite-dimensional  multi-particle HS algebra \cite{multiparticle}  acts on the space of amplitudes. Therefore, higher symmetries of scattering amplitudes of a given QFT, if exist, can be viewed as embedded into the multiparticle HS symmetries. One problem for the future is to apply  the proposed approach to  $\mathcal{N}=4$ SYM.

Conserved multi-particle HS charges are represented as integrals of the $\eta$-dependent on-shell closed differential form \eqref{eq:form}. Since, {\it a priori}, $\eta$ is an arbitrary function of characteristics \eqref{eq:characteristics} giving rise to an infinite set of conserved charges our construction  is likely to be rich enough to embed all possible scattering amplitudes. Since fields $\fld{\rk}{y,\ot{y}}{x}$ \eqref{eq:unfolded} describing massless fields in the unfolded formalism encode all their on-shell nontrivial derivatives, one may expect that space-time non-localities\footnote{\label{note}Note that within the unfolded approach one should distinguish between space-time non-localities and those in the extended space $\mathbf{Cor}$. For example, rational functions of $\lambda_{\ind{i}a}$ and $\ot{\lambda}_{\oind{j}\p{a}}$ are local in $\mathbf{Cor}$, however they actually represent derivatives $\frac{\partial}{\partial y^{a}_{\ind{i}}}$, $\frac{\partial}{\partial \ot{y}^{\p{a}}_{\oind{j}}}$  and hence derivative $\frac{\partial}{\partial x^{a\p{a}}}$ by \eqref{eq:unfolded}, that may result in space-time non-locality.} that may appear for HS charges are sufficient to reproduce  scattering amplitudes represented by rational functions of on-shell momenta $p_{I}$.  This is explicitly checked for the spin-$1$ Parke-Taylor formula \eqref{eq:PTf} and its HS generalization  \eqref{eq:PTf_3hs}. The $n$-point MHV amplitudes with $n\ge 4$ were also obtained with the aid of the space-time non-local rational symmetry parameter \eqref{eq:toMHV_n}.

A related comment is that, as explained in Section \ref{Sec_FormFactors}, instead of working with gauge potentials one can perform computations using gauge invariant field strengths in terms of which the HS conserved charges \eqref{eq:amplitude} were originally constructed. This is achieved by allowing charges \eqref{eq:amplitude} to be non-local in the sense that symmetry parameters $\eta$ are allowed to be rational functions of variables $\lambda_{\ind{i}a}$, $\ot{\lambda}_{\oind{j}\p{a}}$.

A remarkable novel opportunity due to application of the unfolded approach consists in representing the multi-particle amplitudes as integrals of a differential form $\Omega\left[\eta\right]$ \eqref{eq:form}  closed in the correspondence space $\mathbf{Cor}$ ($\dim \mathbf{Cor} > \deg \Omega$) containing both space-time and spinor space as subspaces. To obtain charge \eqref{eq:amplitude} one can take different cycles $\mathbf{S}_1$ and $\mathbf{S}_2$ in the same homotopy class
arriving at different representations of the same conserved charge. The two
representations may be seemingly unrelated unless the total space $\mathbf{Cor}$ with the closed differential form $\Omega$ is brought into play.
This construction is anticipated to provide a proper language for relating different approaches in the amplitude calculations even beyond the HS framework. In this article integration was performed both over the cycle $\mathbf{S}_1$ in the spinor (twistor) space (with coordinates $\lambda_{\ind{i}a},\ot{\lambda}_{\oind{j}\p{a}},y^{a}_{\oind{j}}, \ot{y}^{\p{a}}_{\ind{i}}$ \eqref{eq:amplitude_n} in Section $2$ and $\lambda_{\got{a}}$, $\ot{y}^{\pgot{a}}$ in Section $3$) and over the space-time cycle $\mathbf{S}_2$ (\textit{cf.} \eqref{eq:amplitude_nx}), giving the same result.

Besides usual $4d$ Minkowski space-time $\Mink{2}$ we have also made use of unfolded formulation in the generalized Minkowski space-time $\Mink{K}$ of \cite{higher_rank} which is appropriate for the description of poly-local products of massless fields. Usual Minkowski space is realized as the $4$-dimensional subspace $\Mink{2}\subset\Mink{K}$ via diagonal embedding (see Section \ref{Sec_Mink}). The advantage of this approach  is that it allows us to treat all single-particle legs of the amplitude  in a way symmetric with respect to both spinor and space-time sectors. To distinguish between gluons of opposite helicities in the amplitude one has to take a special solution \eqref{eq:solution_K} of unfolded equation \eqref{eq:unfolded_Fourier_K} which manifestly breaks $\mathfrak{S}_n$ symmetry permuting  single-particle legs. This generalized framework reproduces the $n$-point Parke-Taylor formula as well. Moreover, in the generalized Minkowski space $\Mink{K}$, $n$-particle MHV amplitudes can be interpreted as poly-local charges resulting from  multiple integration over $n$ copies of self-dual two-dimensional planes in the complexified Minkowski space-time \eqref{eq:pullback_Kx}.

 Note that in the interesting recent paper \cite{amplitudes_dif_forms}  scattering amplitudes of a theory were represented in terms of differential forms in momentum variables. As a result integration of differential forms elucidates the underlying geometry and controls the structure of amplitudes via singularities. (Somewhat similar situation took place for the calculation of HS conserved charges in flat space in the framework of \cite{theta} where a $\frac{1}{z}$-singularity in a $\mathbb{C}$-plane had to be inserted  to give a non-zero result.) However what is different between the constructions of \cite{amplitudes_dif_forms} and of this paper is that we  consider on-shell closed differential forms in the correspondence space $\mathbf{Cor}$ which unifies space-time and spinor variables (also including momentum spinors $\lambda_{\ind{i}a}$, $\ot{\lambda}_{\oind{j}\p{a}}$ \eqref{eq:unfolded_Fourier}). This allows us to relate different expressions for a given amplitude via deformation of the integration cycle in $\mathbf{Cor}$. Also, the proposed formalism  is  applicable to amplitudes containing lower and higher spins  while that of \cite{amplitudes_dif_forms} is straightforward only for $s\le 1$.

Apart from application of the proposed technique to the analysis of lower-spin Yang-Mills-type models, it would be interesting to use it for computation of amplitudes resulting from nonlinear HS equations of \cite{non-lin}. This project becomes realistic after the issue of locality in HS equations has been resolved at least in the lowest orders \cite{locality,homotopy,Vas_low_order}. A peculiarity of HS theory is that it is consistently formulated in the $AdS$ space-time \cite{hs_vertices_grav} hence suggesting the computation of flat-space amplitudes starting from $AdS$. This is somewhat reminiscent of the computations of unitary Minkowski amplitudes starting from conformal twistor string in the framework of twistor geometry \cite{Witten,Mc} demanding reduction to the $AdS$ space. Further extension of the proposed construction to string-like HS theories proposed in \cite{Coxeter} is also very interesting.

 Let us stress that our approach is applicable to amplitudes of any kind, 
  {\it i.e.}, not necessarily MHV and/or tree ones. Corrections of all kinds will contribute to the parameter $\eta$ that determines the form of an amplitude. In particular, it would be interesting to extend our results to NMHV amplitudes and loop corrections (see \cite{BCF,Hamed_gras} and references therein), as well as to HS multiparticle amplitudes \cite{BC_hs,Tsulaia_amplitudes,Mc} by reading off the corresponding parameters $\eta$ from the literature.

Among other problems for the future we mention supersymmetrization of the proposed technique to make it more directly related to the supersymmetric models like $\mathcal{N}=4$ SYM.

\section*{Acknowledgements}

We are grateful to Olga Gelfond and Mirian Tsulaia for collaboration at the early stage
of this project and also to Leonid Bork, Ruslan Metsaev, Dmitry Ponomarev and  Alexei Rosly for stimulating discussions. Y.G. thanks the unit ``Physique de l'Univers, Champs et Gravitation'' of UMONS (Belgium) for kind hospitality during the final stage of preparation of this work. This work is supported by RSF grant 18-12-00507.


\begin{thebibliography}{39}

\bibitem{higher_rank}
  O.~A.~Gelfond and M.~A.~Vasiliev,
  %``Higher-Rank Fields and Currents,''
  JHEP {\bf 1610}, 067 (2016)
  %doi:10.1007/JHEP10(2016)067
  [arXiv:1312.6673 [hep-th]].

\bibitem{PT}
S.~Parke and T.~Taylor,
%“An Amplitude For N Gluon Scattering,”
Phys. Rev. Lett. 56 (1986) 2459.

\bibitem{PT_proof}
 F.~A.~Berends and W.~T.~Giele,
 %“Recursive Calculations For Processes With N Gluons,”
 Nucl. Phys. B306 (1988) 759.

\bibitem{Witten}
  E.~Witten,
  %``Perturbative gauge theory as a string theory in twistor space,''
  Commun.\ Math.\ Phys.\  {\bf 252}, 189 (2004)
  %doi:10.1007/s00220-004-1187-3
  [hep-th/0312171].

\bibitem{BCFW}
  R.~Britto, F.~Cachazo, B.~Feng and E.~Witten,
  %``Direct proof of tree-level recursion relation in Yang-Mills theory,''
  Phys.\ Rev.\ Lett.\  {\bf 94}, 181602 (2005)
  %doi:10.1103/PhysRevLett.94.181602
  [hep-th/0501052].

\bibitem{Hamed_S}
  N.~Arkani-Hamed, F.~Cachazo, C.~Cheung and J.~Kaplan,
  %``The S-Matrix in Twistor Space,''
  JHEP {\bf 1003}, 110 (2010)
  %doi:10.1007/JHEP03(2010)110
  [arXiv:0903.2110 [hep-th]].

\bibitem{Hamed_gras}
  N.~Arkani-Hamed, F.~Cachazo, C.~Cheung and J.~Kaplan,
  %``A Duality For The S Matrix,''
  JHEP {\bf 1003}, 020 (2010)
  %doi:10.1007/JHEP03(2010)020
  [arXiv:0907.5418 [hep-th]].

\bibitem{Dual_Super_Conf}
  J.~M.~Drummond, J.~Henn, G.~P.~Korchemsky and E.~Sokatchev,
%  ``Dual superconformal symmetry of scattering amplitudes in N=4 super-Yang-Mills theory,''
  Nucl.\ Phys.\ B {\bf 828}, 317 (2010)
%  doi:10.1016/j.nuclphysb.2009.11.022
  [arXiv:0807.1095 [hep-th]].

\bibitem{Dual_Super_Conf_2}
  A.~Brandhuber, P.~Heslop and G.~Travaglini,
  %``A Note on dual superconformal symmetry of the N=4 super Yang-Mills S-matrix,''
  Phys.\ Rev.\ D {\bf 78}, 125005 (2008)
  %doi:10.1103/PhysRevD.78.125005
  [arXiv:0807.4097 [hep-th]].

\bibitem{Yangian}
 J.~M.~Drummond, J.~M.~Henn and J.~Plefka,
%  ``Yangian symmetry of scattering amplitudes in N=4 super Yang-Mills theory,''
  JHEP {\bf 0905}, 046 (2009)
%  doi:10.1088/1126-6708/2009/05/046
 [arXiv:0902.2987 [hep-th]].

\bibitem{Bork_ff}
  L.~V.~Bork and A.~I.~Onishchenko,
  %``On soft theorems and form factors in $ \mathcal{N}=4 $ SYM theory,''
  JHEP {\bf 1512}, 030 (2015)
  %doi:10.1007/JHEP12(2015)030
  [arXiv:1506.07551 [hep-th]].

\bibitem{form_fact}
  R.~Frassek, D.~Meidinger, D.~Nandan and M.~Wilhelm,
  %``On-shell diagrams, Graßmannians and integrability for form factors,''
  JHEP {\bf 1601}, 182 (2016)
  %doi:10.1007/JHEP01(2016)182
  [arXiv:1506.08192 [hep-th]].

\bibitem{Bork_off_shell}
  L.~V.~Bork and A.~I.~Onishchenko,
  %``Grassmannian integral for general gauge invariant off-shell amplitudes in $ \mathcal{N}=4 $ SYM,''
  JHEP {\bf 1705}, 040 (2017)
  %doi:10.1007/JHEP05(2017)040
  [arXiv:1610.09693 [hep-th]].

\bibitem{perturbiner}
  A.~A.~Rosly and K.~G.~Selivanov,
  %``On amplitudes in selfdual sector of Yang-Mills theory,''
  Phys.\ Lett.\ B {\bf 399}, 135 (1997)
  %doi:10.1016/S0370-2693(97)00268-2
  [hep-th/9611101].

\bibitem{perturbiner_grav}
  A.~A.~Rosly and K.~G.~Selivanov,
  %``Gravitational SD perturbiner,''
  hep-th/9710196.

\bibitem{Mc_actions}
  P.~Hähnel and T.~McLoughlin,
  %``Conformal higher spin theory and twistor space actions,''
  J.\ Phys.\ A {\bf 50}, no. 48, 485401 (2017)
  %doi:10.1088/1751-8121/aa9108
  [arXiv:1604.08209 [hep-th]].

\bibitem{Mc}
  T.~Adamo, P.~Hähnel and T.~McLoughlin,
  %``Conformal higher spin scattering amplitudes from twistor space,''
  JHEP {\bf 1704}, 021 (2017)
  %doi:10.1007/JHEP04(2017)021
  [arXiv:1611.06200 [hep-th]].

%\cite{Shaynkman:2001ip}
\bibitem{Shaynkman:2001ip}
  O.~V.~Shaynkman and M.~A.~Vasiliev,
  %``Higher spin conformal symmetry for matter fields in (2+1)-dimensions,''
  Theor.\ Math.\ Phys.\  {\bf 128} (2001) 1155
   [Teor.\ Mat.\ Fiz.\  {\bf 128} (2001) 378]
 % doi:10.1023/A:1012399417069
  [hep-th/0103208].

%\cite{Vasiliev:2001zy}
\bibitem{Vasiliev:2001zy}
  M.~A.~Vasiliev,
  %``Conformal higher spin symmetries of 4-d massless supermultiplets and osp(L,2M) invariant equations in generalized (super)space,''
  Phys.\ Rev.\ D {\bf 66} (2002) 066006
%  doi:10.1103/PhysRevD.66.066006
  [hep-th/0106149].

\bibitem{theta}
  O.~A.~Gelfond and M.~A.~Vasiliev,
  %``Higher Spin Fields in Siegel Space, Currents and Theta Functions,''
  JHEP {\bf 0903}, 125 (2009)
  %doi:10.1088/1126-6708/2009/03/125
  [arXiv:0801.2191 [hep-th]].

\bibitem{BRST}
  O.~A.~Gelfond and M.~A.~Vasiliev,
  %``Sp(8) invariant higher spin theory, twistors and geometric BRST formulation of unfolded field equations,''
  JHEP {\bf 0912}, 021 (2009)
 % doi:10.1088/1126-6708/2009/12/021
  [arXiv:0901.2176 [hep-th]].

\bibitem{twistors}
O.~A.~Gelfond and M.~A.~Vasiliev,
  %``Operator algebra of free conformal currents via twistors,''
  Nucl.\ Phys.\ B {\bf 876}, 871 (2013)
  %doi:10.1016/j.nuclphysb.2013.09.001
  [arXiv:1301.3123 [hep-th]].

\bibitem{multiparticle}
  M.~A.~Vasiliev,
  %``Multiparticle extension of the higher-spin algebra,''
  Class.\ Quant.\ Grav.\  {\bf 30}, 104006 (2013)
  %doi:10.1088/0264-9381/30/10/104006
  [arXiv:1212.6071 [hep-th]].

\bibitem{Coxeter}
  M.~A.~Vasiliev,
  %``From Coxeter Higher-Spin Theories to Strings and Tensor Models,''
  JHEP {\bf 1808} (2018) 051
 % doi:10.1007/JHEP08(2018)051
  [arXiv:1804.06520 [hep-th]].

\bibitem{chargesAdS}
O.~A.~Gelfond and M.~A.~Vasiliev,
  %``Conserved higher-spin charges in $AdS_4$,''
  Phys.\ Lett.\ B {\bf 754}, 187 (2016)
  %doi:10.1016/j.physletb.2016.01.018
  [arXiv:1412.7147 [hep-th]].

\bibitem{BC_hs}
  P.~Benincasa and F.~Cachazo,
  %``Consistency Conditions on the S-Matrix of Massless Particles,''
  arXiv:0705.4305 [hep-th].

\bibitem{Hamed_hs}
  N.~Arkani-Hamed, T.~C.~Huang and Y.~t.~Huang,
  %``Scattering Amplitudes For All Masses and Spins,''
  arXiv:1709.04891 [hep-th].

%\cite{Vasiliev:1988sa}
\bibitem{Vasiliev:1988sa}
  M.~A.~Vasiliev,
  %``Consistent Equations for Interacting Massless Fields of All Spins in the First Order in Curvatures,''
  Annals Phys.\  {\bf 190} (1989) 59.
 % doi:10.1016/0003-4916(89)90261-3

\bibitem{vasiliev_starprod_ads}
 M.~A.~Vasiliev,
  %``Higher spin gauge theories: Star product and AdS space,''
  In *Shifman, M.A. (ed.): The many faces of the superworld* 533-610
  %doi:10.1142/9789812793850\_0030
  [hep-th/9910096].

\bibitem{Sorokin_hyperspace}
  D.~Sorokin and M.~Tsulaia,
  %``Higher Spin Fields in Hyperspace. A Review,''
  Universe {\bf 4}, no. 1, 7 (2018)
  %doi:10.3390/universe4010007
  [arXiv:1710.08244 [hep-th]].

%\cite{Gelfond:2003vh}
\bibitem{Gelfond:2003vh}
  O.~A.~Gelfond and M.~A.~Vasiliev,
  %``Higher rank conformal fields in the Sp(2M) symmetric generalized space-time,''
  Theor.\ Math.\ Phys.\  {\bf 145} (2005) 1400
   [Teor.\ Mat.\ Fiz.\  {\bf 145} (2005) 35]
  %doi:10.1007/s11232-005-0168-9
  [hep-th/0304020].

\bibitem{Gelf_Shil}
  I.~M.~Gelfand, G.~E.~Shilov,
  ``Generalized functions. Volume 1. Properties and operations,''
  Academic Press, New York and London, 1964.

\bibitem{GY}
  S.~Giombi and X.~Yin,
  %``The Higher Spin/Vector Model Duality,''
  J.\ Phys.\ A {\bf 46}, 214003 (2013)
  %doi:10.1088/1751-8113/46/21/214003
  [arXiv:1208.4036 [hep-th]].

\bibitem{locality}
  V.~E.~Didenko and M.~A.~Vasiliev,
  %``Test of the local form of higher-spin equations via AdS / CFT,''
  Phys.\ Lett.\ B {\bf 775}, 352 (2017)
  %doi:10.1016/j.physletb.2017.09.091
  [arXiv:1705.03440 [hep-th]].

\bibitem{homotopy}
  O.~A.~Gelfond and M.~A.~Vasiliev,
  %``Homotopy Operators and Locality Theorems in Higher-Spin Equations,''
  Phys.\ Lett.\ B {\bf 786} (2018) 180
  [arXiv:1805.11941 [hep-th]].

\bibitem{BBB}
  A.~K.~H.~Bengtsson, I.~Bengtsson and L.~Brink,
  %``Cubic Interaction Terms for Arbitrary Spin,''
  Nucl.\ Phys.\ B {\bf 227}, 31 (1983).
  %doi:10.1016/0550-3213(83)90140-2

\bibitem{BBL}
  A.~K.~H.~Bengtsson, I.~Bengtsson and N.~Linden,
  %``Interacting Higher Spin Gauge Fields on the Light Front,''
  Class.\ Quant.\ Grav.\  {\bf 4}, 1333 (1987).
 % doi:10.1088/0264-9381/4/5/028

\bibitem{FraMets}
  E.~S.~Fradkin and R.~R.~Metsaev,
  %``Cubic scattering amplitudes for all massless representations of the Poincare group in any space-time dimension,''
  Phys.\ Rev.\ D {\bf 52}, 4660 (1995).
  %doi:10.1103/PhysRevD.52.4660

\bibitem{Coleman_Mandula}
  S.~R.~Coleman and J.~Mandula,
  %``All Possible Symmetries of the S Matrix,''
  Phys.\ Rev.\  {\bf 159}, 1251 (1967).
  %doi:10.1103/PhysRev.159.1251

\bibitem{Mar} M.~A.~Vasiliev,
%`` Relativity, Causality, Locality, Quantization and Duality
%in the $Sp(2M)$ Invariant Generalized Space-Time'',
{\tt hep-th/0111119};
 Contribution to the Marinov's Memorial
Volume, M.Olshanetsky and A.Vainshtein Eds, World Scientific.

%\cite{Bandos:2005mb}
\bibitem{Bandos:2005mb}
  I.~Bandos, X.~Bekaert, J.~A.~de Azcarraga, D.~Sorokin and M.~Tsulaia,
  %``Dynamics of higher spin fields and tensorial space,''
  JHEP {\bf 0505} (2005) 031
  %doi:10.1088/1126-6708/2005/05/031
  [hep-th/0501113].

\bibitem{amplitudes_dif_forms}
  S.~He and C.~Zhang,
  %``Notes on Scattering Amplitudes as Differential Forms,''
  arXiv:1807.11051 [hep-th].

\bibitem{non-lin}
  M.~A.~Vasiliev,
 %``More on equations of motion for interacting massless fields of all spins in (3+1)-dimensions,''
  Phys.\ Lett.\ B {\bf 285}, 225 (1992).
  %doi:10.1016/0370-2693(92)91457-K

\bibitem{Vas_low_order}
 V.~E.~Didenko, O.~A.~Gelfond, A.~V.~Korybut and M.~A.~Vasiliev,
  %``Homotopy Properties and Lower-Order Vertices in Higher-Spin Equations,''
  J.\ Phys.\ A {\bf 51} (2018) no.46,  465202
 % doi:10.1088/1751-8121/aae5e1
  [arXiv:1807.00001 [hep-th]].


\bibitem{hs_vertices_grav}
  E.~S.~Fradkin and M.~A.~Vasiliev,
  %``On the Gravitational Interaction of Massless Higher Spin Fields,''
  Phys.\ Lett.\ B {\bf 189}, 89 (1987).

\bibitem{BCF}
  R.~Britto, F.~Cachazo and B.~Feng,
  %``New recursion relations for tree amplitudes of gluons,''
  Nucl.\ Phys.\ B {\bf 715}, 499 (2005)
  %doi:10.1016/j.nuclphysb.2005.02.030
  [hep-th/0412308].

\bibitem{Tsulaia_amplitudes}
  P.~Dempster and M.~Tsulaia,
  %``On the Structure of Quartic Vertices for Massless Higher Spin Fields on Minkowski Background,''
  Nucl.\ Phys.\ B {\bf 865}, 353 (2012)
  %doi:10.1016/j.nuclphysb.2012.07.031
  [arXiv:1203.5597 [hep-th]].

\end{thebibliography}
\end{document}